\documentclass[aps,prb,floatfix,amsmath,amssymb,preprint,eqsecnum,nofootinbib]{revtex4}
\usepackage{graphicx}
\usepackage{dcolumn}
\usepackage{epstopdf}
\usepackage{color}

\usepackage{setspace} 
\newcommand{\ea}{{{\bf e}_1}}
\newcommand{\eb}{{{\bf e}_2}}
\newcommand{\ec}{{{\bf e}_3}}

\newcommand{\beq}{\begin{equation}}
\newcommand{\eeq}{\end{equation}}
\newcommand{\beqn}{\begin{eqnarray}}
\newcommand{\eeqn}{\end{eqnarray}}
\newcommand{\nn}{\nonumber\\}
\newcommand{\bk}{{\bf k}}
\newcommand{\bq}{{\bf q}}
\newcommand{\bp}{{\bf p}}
\newcommand{\br}{{\bf r}}
\newcommand{\bQ}{{\bf Q}}
\newcommand{\bm}{{m^a}}
\newcommand{\bn}{n^a}
\newcommand{\bs}{{\sigma^a}}
\newcommand{\bx}{{\bf x}}

\newcommand{\bA}{{\mathcal{A}^{\mu}}}

\usepackage{amsmath}
\usepackage{amssymb}
\begin{document}

\title{Geometric phases and competing orders in two dimensions}

\author{Liang Fu}
\affiliation{Department of Physics, Harvard University, Cambridge MA
02138}

\author{Subir Sachdev}
\affiliation{Department of Physics, Harvard University, Cambridge MA
02138}

\author{Cenke Xu}
\affiliation{Department of Physics, University of California, Santa Barbara CA 93106}

\date{\today \\
\vspace{1.6in}}

\begin{abstract}
We discuss the problem of characterizing ``quantum disordered'' ground states, obtained
upon loss of antiferromagnetic order on general lattices in two spatial dimensions, with arbitrary electronic
band structure. A key result is the response in electron bilinears to the skyrmion
density in the local antiferromagnetic order, induced by geometric phases.
We also discuss the connection to topological terms obtained under situations
where the electronic spectrum has a Dirac form.
\end{abstract}

\maketitle

\section{Introduction}
\label{sec:intro}

Geometric phases have played a central role in many fields of
physics, and especially in the quantum Hall effect at high
magnetic fields\cite{shapere}. However, in subsequent research it
has also become clear that geometric phases are crucial for a
complete understanding of the quantum phase transitions of
correlated electron systems in zero applied magnetic field.
Traditionally, classical phase transitions are described in terms
of an `order parameter', with one phase being ordered, and the
other `disordered'. Upon extending this idea to quantum phase
transitions, we have the possibility of a `quantum-disordered'
phase \cite{chn}. However, in almost all of the interesting examples,
the latter phase is not disordered: geometric phases induce
a `competing' order. A separate possibility is that the
quantum-disordered phase has fractionalization and topological order:
we will not explore this latter possibility in the present
paper.

(We note here that the word `phase' has two separate meanings
above, and in the remainder of the paper. When used by itself,
`phase' refers to a particular state of a thermodynamic system.
However, in the combination `geometric phase', it refers to the
angular co-ordinate of a complex number representing the
wavefunction. We trust the context will clarify the meaning for
the reader.)

In two dimensional systems, the earliest example of a competing
order induced by geometric phases was in the spin $S=1/2$ square
lattice Heisenberg antiferromagnet. The model with
nearest-neighbor interactions has long-range N\'eel order. We can
try to destroy this order by adding further neighbor frustrating
interactions, leading to a possible quantum-disordered phase
\cite{chn}. Such a phase should be characterized by the
proliferation of defects in the N\'eel order: for collinear N\'eel
order in a model with SU(2) spin symmetry, the order parameter
lies on $S^2$ (the surface of a sphere), and the homotopy group
$\pi_2 (S^2) = Z$, implies that the existence of point defects
known as hedgehogs. Haldane \cite{haldane} pointed out that
geometric phases of the lattice spins endowed each hedgehog with a
net geometric phase, and argued that this implied a 4-fold
degeneracy of the quantum-disordered ground state. Read and
Sachdev \cite{rs1,rs2} demonstrated that the hedgehog geometric
phase actually implied a competing order, associated with a broken
lattice symmetry due to valence bond solid (VBS) order. The VBS
order can take 4 orientations related by lattice symmetries, thus
realizing the 4-fold degeneracy. They also presented two
additional derivations of the hedgehog geometric phase: from the
Schwinger boson representation of the spins \cite{rs2}, and via a
duality transform of the quantum dimer model \cite{rk}.
Sachdev and Jalabert \cite{jalabert} introduced a lattice gauge theory for
the competing N\'eel and VBS orders, in which the geometric phase
appeared as a coupling between the skyrmion density associated with the N\'eel order
(defined in Section~\ref{sec:neel}) and a lattice field linked to the VBS order.

We note in passing that we will not be interested here on the
separate question of the nature of the phase transition between
the two competing order phases. A second order transition appears
in the `deconfined criticality' theory \cite{senthil1,senthil2},
and this proposal has been the focus of a number of numerical
studies
\cite{sandvik,melkokaul,wiese,kuklov,mv,sandvik2,sandvik3,damle}.

A different perspective on the N\'eel-VBS transition appeared in the
work of Tanaka and Hu \cite{tanakahu}
who used a continuum theory of Dirac fermions.
The previous works had all represented the spins in terms of
bosonic degrees of freedom which carried geometric phases. Tanaka
and Hu instead used a fermionic representation of the spins, and
then chose a low-energy limit which allowed representation of the
theory in terms of continuum Dirac fermions in 2+1 spacetime
dimensions. The Dirac representation appeared from a band
structure of the lattice fermions in which there was $\pi$ flux
per plaquette: this could be interpreted as the dispersion of
fermionic spinons in a a particular algebraic spin liquid (ASL)
important for intermediate length-scale physics, or as a
mean-field dispersion of electrons in a particular extended
Hubbard model \cite{senthilfisher}. Armed with the Dirac fermions,
Tanaka and Hu used
field-theoretic developments by Abanov and Wiegmann \cite{abanov}
to show that the effective action for the N\'eel and
VBS order parameters allowed representation of the geometric phase
as a Wess-Zumino-Witten (WZW) term for a 5-component order
parameter. The co-efficient of this WZW term was quantized to a value
reduced consistently to the hedgehog Berry phases in the appropriate limit.
In a different context, Grover and Senthil
\cite{groversenthil} recently showed that a WZW term was also
present between a quantum spin Hall order parameter and $s$-wave
superconducting order on the honeycomb lattice; their computation also
used the Dirac spectrum of the electrons on the honeycomb lattice.

The appearance of the WZW term with quantized co-efficient in the above computation
appears surprising from the perspective of the earlier bosonic
formulations \cite{haldane,rs1,rs2}. In these earlier works, the quantization was directly
related to the quantization of the spin on each lattice site, which relied crucially
on the projection to one electron (in general, to $2S$ electrons in a fully antisymmetric orbital state, for spin $S$)
on every site. In contrast, in the above fermionic formulations \cite{tanakahu,senthilfisher,groversenthil},
the local constraints are ignored in the computation of the WZW term, apart from a global constraint
on the average fermion density. We will argue in this paper that
the WZW term with a quantized co-efficient is an artifact of the low energy Dirac fermionic spectrum.

This aim of our paper is extend the use of the fermionic representation
of the underlying degrees of freedom to cases without a low energy Dirac limit.
We will develop a general approach to computing geometric phases,
which works for
arbitrary electronic band structures, whether insulating, metallic or superconducting.
Like the recent work \cite{tanakahu,senthilfisher,groversenthil}, we will not impose
a local constraint on the electron number, which is permissible for the
metallic or superconducting cases or even in insulators with small on-site repulsive energy.
Our computation begins by applying local antiferromagnetic order, and computing
the band structure in the presence of this order. Then, we allow the orientation of the local
order to become spacetime-dependent, so that eventually there is no true
long-range antiferromagnetic order. However, the local ordering is still assumed
to be present, with its associated band structure, and we fill these electronic states up to the
Fermi level. We will then compute response of these
filled electronic states to spatial variations in the antiferromagnetic order.
We will also allow spatial variations in competing orders, deduce their coupling
to antiferromagnetism. We will find geometric phases between the order parameters,
but will show that a WZW representation does not exist in general.

Another approach to the general problem of geometric phases was
described recently by Yao and Lee \cite{yaolee}. Their method
required extension \cite{qi} of the 2 dimensional electronic band structure
to 6 dimensions, and the computation of topological invariants in
6 dimensions and of the mapping between 2 and 6 dimensions.
Non-zero values of these invariants were then argued to be
sufficient conditions for a WZW term in the effective action for
the competing orders. This last conclusion appears to be at variance with
our results.

We begin in Section~\ref{sec:neel} by considering spatial
variations in the antiferromagnetic order on the square lattice.
We compute the response to this spatially varying background,
in the spirit of the computation of Chern
numbers of integer quantum Hall states by Thouless {\em et al.}
\cite{tkkn}. This leads to the key result in Eq.~(\ref{omain}).

Section~\ref{sec:liang} extends the computation to allow for
simultaneous variation of both N\'eel and VBS orders.
Here we will also make a connection to the dimensional reduction
method \cite{qi,yaolee} noted above.
Section~\ref{sec:honeycomb} contains applications of our results
to insulators on the honeycomb lattice, while Section~\ref{sec:nodal}
considers transitions in the background of the nodal quasiparticles
of a $d$-wave superconductor.

\section{Fluctuating N\'eel order}
\label{sec:neel}

Our approach begins with with an arbitrary band structure for
lattice fermions $c_{\alpha}$, with the spin index $\alpha =
\uparrow, \downarrow$; so the band structure of the electronic quasiparticles
is \beq H_b = -\sum_{i,j} t( {\br}_i - {\br}_j) c^\dagger_\alpha
({\br}_i) c_\alpha ({\br}_j) \label{e1} \eeq where $\br_i$ labels
the lattice sites, and $t (\br)$ are the tight-binding hopping
matrix elements. For definiteness, let us consider the N\'eel
state on the square lattice, as described by the Slater mean-field
theory of antiferromagnetic order. We allow the N\'eel order to
have a slow spatial variation in its orientation, which we specify
by the unit vector $n^a (\br)$ ($a = x,y,z$). In this modulated
N\'eel state, the electronic quasiparticle Hamiltonian is modified
from the band structure in Eq.~(\ref{e1}) to
\beq
H = -\sum_{i,j} t( {\bf r}_i - {\bf
r}_j) c_\alpha^\dagger ({\bf r}_i) c_\alpha ({\bf r}_j) + m
\sum_i \eta_i n^a (\br_i ) c_\alpha^\dagger ({\bf r}_i)
\sigma^a_{\alpha\beta} c_\beta ({\bf r}_i) \label{e2}
\eeq
where
$\sigma^a$ are the spin Pauli matrices, $\eta_i = \pm 1$ on the
two sublattices of the N\'eel order, and $m$ is a mean-field
magnitude of the band splitting due to the N\'eel order. The main
result of the following Section~\ref{sec:linear} will be
obtained by working directly with Eq.~(\ref{e2}) for a slow
variation of $n^a (\br)$ about a fully polarized N\'eel state.

For some purposes, we will find it advantageous to use an alternative
gauge-theoretic formulation, which has some technical advantages
for a global perspective on the phase diagram. For this, we follow
Ref.~\onlinecite{su2}, and transform to a rotating reference frame
in the varying N\'eel background so that the N\'eel order points
in the constant direction (0,0,1) in the new reference frame. We
do this by introducing complex bosonic spinors $z_{i \alpha}$,
with $|z_{i \uparrow}|^2 + |z_{i \downarrow}|^2 = 1$ so that
\beq
\left( \begin{array}{c} c_{ \uparrow} \\ c_{ \downarrow}
\end{array} \right) =
 \left( \begin{array}{cc} z_{\uparrow} & -z_{\downarrow}^\ast \\
z_{\downarrow} & z_{\uparrow}^\ast \end{array} \right) \left(
\begin{array}{c} \psi_{ +} \\ \psi_{ -} \end{array} \right)
\label{zpsi}
\eeq
where $\psi_{p}$, $p = \pm$, are the
``electrons'' in the rotating reference frame. We will assume that
the $z_{ \alpha}$ have a slow dependence upon spacetime, allowing
in expansion in gradients of the $z_{\alpha}$. A fixed orientation
of the N\'eel order is realized in the rotating reference frame by
choosing the $z_\alpha$ so that
\beq
n^a = z_\alpha^\ast
\sigma^a_{\alpha\beta} z_\beta \label{nz}
\eeq
However, we will
{\em not} assume any slow variations in the fermions $c_{\alpha}$
and $\psi_{p}$, allowing them to carry arbitrary momenta and band
structures.

Parameterizations like (\ref{zpsi}) were motivated earlier by the
Schwinger boson formulation of the underlying antiferromagnet. In
such theories, the geometric phases of the spins at half-filling
were associated entirely with those of the Schwinger bosons
\cite{rs2}. In our computations of geometric phases in the present
paper, we will find it convenient to work in an approach in which
the lattice geometric phases are attached entirely to fermionic
degrees of freedom. For this, we will use an exact rotor model
formulation of a general lattice Hamiltonian for which
Eq.~(\ref{zpsi}) also holds. The details of this rotor formulation
are presented in Appendix~\ref{app:rotor}, and this should be
regarded as an alternative to earlier Schwinger boson
formulations. In the rotor theory, the $\psi_\pm$ are canonical
fermions with a density equal to the full electron density; thus
in the the insulator, the total $\psi_\pm$ density is 1, and it is
this unit density which leads to the geometric phases. The bosonic
variables $z_\alpha$ have a rotor kinetic energy with only a
second-order time-derivative in the action {\em i.e.\/} they are
not canonical bosons, and do not directly carry any geometric
phases. In the Schwinger boson formulation, the bosons are
canonical, and this complicates the computation of geometric
phases in the general case.

Inserting Eq.~(\ref{zpsi}) into Eq.~(\ref{e2}), we obtain the
theory for the $\psi_\pm$ fermions, which we write in the form
\cite{su2}
\beq
H = -\sum_{i,j} t( {\bf r}_i - {\bf r}_j)
\psi_p^\dagger ({\bf r}_i) e^{i p A_{ij}} \psi_p ({\bf r}_j) + m
\sum_i \eta_i \, p \, \psi_p^\dagger ({\bf r}_i)  \psi_p ({\bf
r}_i)  + \ldots \label{e3}
\eeq
First, note that the transformation
to the rotating reference frame has removed the slowly varying $r$
dependence from the second term proportional to $m$. Instead the
effect of the transformation into the rotating reference is now
entirely in the hopping term. As discussed in earlier work
\cite{su2}, these modifications can be expressed in general in
terms of a SU(2) gauge potential, corresponding to the SU(2) gauge
redundancy introduced by the parameterization in Eq.~(\ref{zpsi}).
In the fluctuating N\'eel state we consider here, the SU(2) gauge
invariance is `Higgsed' down to U(1): this corresponds to the
invariance of Eq.~(\ref{nz}) only under a U(1) gauge
transformation of the $z_\alpha$. So we write only the U(1) gauge
potential term in Eq.~(\ref{e3}), represented by $A_{ij}$.
The ellipses in Eq.~(\ref{e3}) refer to additional fermion hopping terms
connected to the remaining SU(2) gauge fields: these were written
out explicitly in Refs.~\onlinecite{su2,yang}, and also appear in the present
paper as the last two terms in Eq.~(\ref{hop}).

As we are using a continuum formulation for the order parameter
$n^a (\br )$ and the $z_\alpha$, we should also work with a
continuum U(1)  gauge potential ${\bf A} (\br)$.
This is related to $A_{ij}$ by an
integral on straight line between $\br_i$ and $\br_j$
\beq
A_{ij}
=  \int_0^1 du \, {\bf A} \left({\bf r}_i + u ({\bf r}_j - {\bf
r}_i) \right) \cdot  ({\bf r}_j - {\bf r}_i) \label{Aij}
\eeq
The flux in the continuum gauge field ${\bf A}$ can be related to
the `skyrmion density' in the antiferromagnetic order parameter:
\beq
\partial_x A_y - \partial_y A_x = \frac{1}{2} \epsilon_{a b c} n^a \partial_x n^b \partial_y n^c . \label{gaugeskyrmion}
\eeq
With periodic boundary conditions, the spatial integral of the skyrmion density on the right-hand-side is
a topological invariant, and is quantized to an integer multiple of $2 \pi$; the integer is the skyrmion number.
Thus inducing a $2 \pi$ flux in ${\bf A}$ corresponds to
changing the skyrmion number of the field $n^a (\br )$ by unity,
which is the same as introducing a hedgehog defect in the N\'eel order.

\subsection{Response to spin textures}
\label{sec:linear}

This section will carry out the formally simple exercise of
computing the linear response of the Hamiltonian Eq.~(\ref{e2})
a slowly varying spacetime dependence in the order parameter $n^a (\br )$.
A similar computation can also be carried out using the alternative
gauge-theoretic form in Eq.~(\ref{e3}) to a slowly varying gauge potential
$A_{ij}$: the latter computation is presented in Appendix~\ref{app:gauge}.

We begin with the Hamiltonian in Eq.~(\ref{e2}), and assume $n^a
(\br )$ is a slowly varying unit vector. In any local region, without loss
of generality, we can choose
co-ordinates so that $n^a (\br )$ is close to the North pole $(0,0,1)$.
In this co-ordinate system, as in
Ref.~\onlinecite{yang}, we parameterize the
variations in the N\'eel order in
terms of
the complex field $\varphi$ via
\beq
n^a = \left(  \frac{\varphi +
\varphi^\ast}{2},  \frac{\varphi -
\varphi^\ast}{2i}, \sqrt{1 -  |\varphi|^2} \right). \label{nphi}
\eeq
We assume $|\varphi| \ll 1$ and slowly varying.
Inserting Eq.~(\ref{nphi}) into Eq.~(\ref{e2}) we obtain the
Hamiltonian $H=H_0 + H_1$ with
\beqn H_0 &=& \sum_{{\bk}} \left( \varepsilon_{{\bk}}
c^{\dagger} ({\bk}) c ({\bk}) + m c^\dagger ({\bk} +
\bQ) \sigma^z c ({\bk}) \right),
\label{eh0}
\eeqn
where $\bQ = (\pi, \pi)$
and \beq \varepsilon_\bk = -\sum_{{\bf s}} t({\bf s}) \cos({\bk}
\cdot {\bf s}), \eeq with $t(-{\bf s}) = t({\bf s})$. Throughout
this section, the summation over momenta extends over the entire
square lattice Brillouin zone. Also, we will drop the $\alpha$
spin indices of the $c_\alpha$, all Pauli matrices in this present
section will be assumed to act on the $\alpha$ space, and the $\alpha$ indices will be traced
over.
The coupling to the spatial variations in the N\'eel order parameterized
by $\varphi$ are given to the needed order in $\varphi$ by
\beqn
H_1 &=&  m \sum_{\bk_1, \bk_2} \Bigl[ \varphi^\ast (\bk_1 ) c^\dagger (\bk_2 +
\bQ) \sigma^+ c (\bk_2 + \bk_1) + \varphi (\bk_1 ) c^\dagger (\bk_2 +
\bQ) \sigma^- c (\bk_2 - \bk_1) \Bigr] \nn
&~& \quad - \frac{m}{2} \sum_{\bk_1, \bk_2, \bk_3} \varphi^\ast ( \bk_1 ) \varphi (\bk_1 + \bk_2)  c^\dagger (\bk_3 +
\bQ) \sigma^z c (\bk_3 - \bk_1) \label{a2}
\eeqn

We are now interesting in computing the response of
the observable properties of $H$ to a slow variation in the N\'eel order
$n^a ( \br )$. A key choice we have to make here is that of a suitable
observable. We are interested in the nature of the phase where N\'eel order
is `disordered' and so it is natural that the observable should be spin rotation
invariant. Also, because we will use the observable to characterize a `competing order',
it should preferably vanish in the spatially uniform N\'eel state, and be induced
only when there are spatial variations in the N\'eel order. Finally, for convenience, the observable
should be a fermion bilinear. With these constraints, it turns out that a
unique choice is forced upon us: it is the observable
\beq
\mathcal{O} (\bk, \br) = \int_q \left\langle c^\dagger (\bk + \bQ + \bq/2) c(\bk - \bq/2) \right\rangle e^{- i \bq \cdot \br}
\eeq
Here the integral over $\bq $ is over {\em small\/} momenta, characteristic of those carried by the bosonic
fields; thus the variation of $\mathcal{O} (\bk, \br)$ with $\br$ is slow. In the simplest case, the right-hand-side
has support only at $\bq=0$,  so that $\mathcal{O} (\bk, \br)$ takes the $\br$-independent
value
\beq
\mathcal{O} (\bk) = \left\langle c^\dagger (\bk + \bQ ) c(\bk ) \right\rangle .
\eeq
 On the other hand,
$\bk$ is an arbitrary momentum in the Brillouin zone, and we will find very useful
information in the $\bk$ dependence of $\mathcal{O} (\bk)$. It is easy to check from $H_0$ that
$\mathcal{O} (\bk ) = 0$ in the uniform N\'eel state, as we required; only
$\left\langle c^\dagger (\bk + \bQ) \sigma^z c (\bk) \right\rangle \neq 0$ in the uniform N\'eel state.
We present an alternative derivation of the choice of the observable $\mathcal{O}$ in Appendix~\ref{app:gauge}:
there we consider an arbitrary fermion bilinear, and show that it is $\mathcal{O}$ which is uniquely
induced to leading order in the applied gauge flux.

We now proceed to a computation of $\mathcal{O} (\bk, \br)$ in powers of $\varphi$ using
the Hamiltonian $H_0 + H_1$. We will need to work to second order in $\varphi$, and also
to second order in spatial gradients of $\varphi$; as stated earlier, all fermion momenta are allowed
to be arbitrary at all stages.

First, let us collect the propagators of $H_0$.
The single
fermion Green's function of $H_0$ is written in terms of its `normal' and
`anomalous' parts as
\beqn \langle c (\bk) \, ;
\, c^{\dagger} (\bp) \rangle &=& \delta_{\bk,\bp} G (\bk ) +
\delta_{\bk+\bQ,\bp} \sigma^z  F (\bk ) \nn
G (\bk ) &\equiv& \frac{u_\bk^2}{-i \omega + E_{1 \bk}} + \frac{v_\bk^2}{-i \omega +
E_{2 \bk}} \nn
F (\bk ) &\equiv&
u_\bk
v_\bk \left( \frac{1}{-i \omega + E_{1 \bk}} - \frac{1}{-i \omega
+ E_{2 \bk}} \right),
\label{green}
\eeqn
where $\bk$ takes all
values in the square lattice Brillouin zone. The eigenenergies in
Eq.~(\ref{green}) are \beq E_{1,2 \bk} = \frac{\varepsilon_\bk +
\varepsilon_{\bk + \bQ}}{2} \pm \sqrt{ \left(\frac{\varepsilon_\bk
- \varepsilon_{\bk + \bQ}}{2}\right)^2 + m^2}, \label{E12} \eeq and the
parameters are \beq u_\bk =  \cos (\theta_\bk/2) \quad, \quad
v_\bk = \sin (\theta_\bk/2) \eeq with \beq \tan \theta_\bk =
\frac{m}{(\varepsilon_\bk - \varepsilon_{\bk + \bQ})/2} \quad ,
\quad 0 < \theta_\bk < \pi \eeq Note that these relations imply
\beq u_{\bk + \bQ} = v_\bk \quad , \quad v_{\bk + \bQ} = u_\bk
\quad, \quad E_{1, \bk + \bQ} = E_{1 \bk} \quad, \quad
 E_{2, \bk + \bQ} = E_{2 \bk}. \eeq

\begin{figure}
\includegraphics[width=2in]{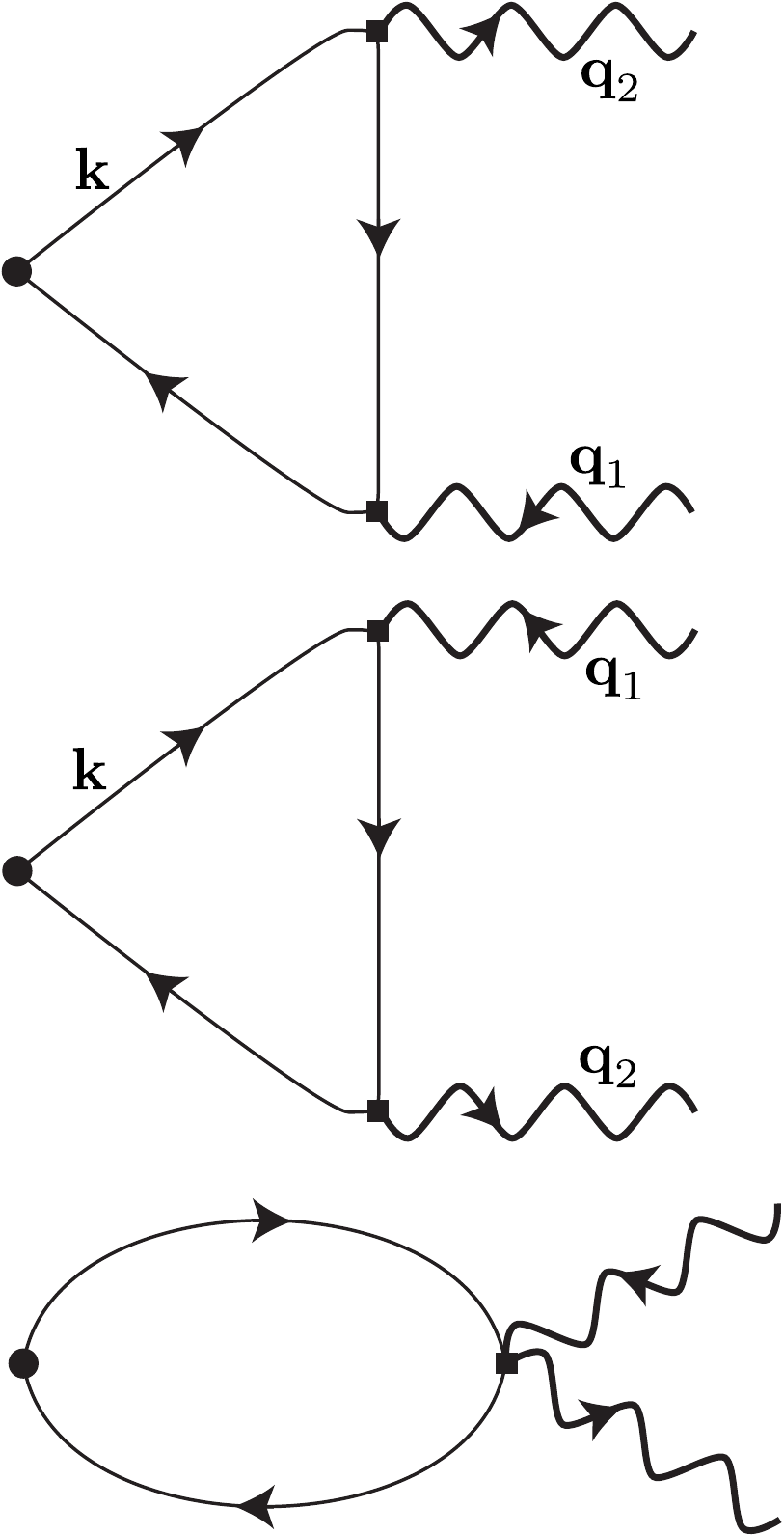}
\caption{Diagrammatic perturbation theory for $\mathcal{O}$
using the couplings in $H_1$ in Eq.~(\ref{a2}).
The wavy lines are $\varphi$ sources, the filled circle is the $\mathcal{O}$ source,
while the full lines are $c$ propagators.}
\label{fig:spinwave}
\end{figure}

The contributions to $\langle c^\dagger (\bk + \bQ + \bq/2) c (\bk -\bq/2) \rangle$ to second order in
$\varphi$ are shown in Fig.~\ref{fig:spinwave}.
The last diagram vanishes identically, while the first two evaluate to
\beq
\left\langle c^\dagger (\bk + \bQ) c (\bk + \bq_1 - \bq_2) \right\rangle =
\sum_{\bq_1, \bq_2} J(\bk, \bq_1, \bq_2) \varphi^\ast (\bq_2) \varphi
(\bq_1 ) \label{J1}
\eeq
where
\beqn
J(\bk, \bq_1, \bq_2) &=&
 m^2 \sum_\omega \Biggl[
F (\bk) G (\bk + \bQ - \bq_2 ) G (\bk + \bq_1 - \bq_2 ) \nn
&-&
G (\bk + \bQ) F (\bk  - \bq_2 ) G (\bk + \bq_1 - \bq_2 ) +
G (\bk + \bQ) G (\bk  - \bq_2 ) F (\bk + \bq_1 - \bq_2 ) \nn
&-&
F (\bk ) F (\bk  - \bq_2 ) F (\bk + \bq_1 - \bq_2 ) \Biggr] - (\bq_1 \leftrightarrow - \bq_2)
\eeqn
We now expand this to second order in  $\bq_1$ and $\bq_2$. This leads to very lengthy
expressions, which we simplified using Mathematica. In the end, a simple final result
was obtained:
\beq
J (\bk, \bq_1, \bq_2) = (\bq_1 \times \bq_2)  \left( \frac{\partial
\varepsilon_{\bk+\bQ}}{\partial \bk} \times \frac{\partial
\varepsilon_\bk}{\partial \bk} \right) \sum_\omega \frac{ m^3}{(-i \omega + E_{1\bk})^3 (- i \omega + E_{2\bk})^3}
\label{J2}
\eeq
Now we combine Eqs.~(\ref{J1}) and (\ref{J2}). The Fourier transform of
$(\bq_1 \times \bq_2)\varphi^\ast (\bq_2) \varphi
(\bq_1 )$ is $\partial_x \varphi \partial_y \varphi^\ast - \partial_y \varphi \partial_x \varphi^\ast$ and to
second order in $\varphi$ this equals $-2i(\partial_x n^x \partial_y n^y - \partial_x n^y \partial_y n^x)$.
In a spin rotationally invariant form, this expression is proportional to the skyrmion density, and so
we have one of our main results:
\beq
\mathcal{O} (\bk, \br) = -i \mathcal{F} (\bk) \epsilon_{abc} n^a (\br) \partial_x n^b (\br) \partial_y n^c (\br) \label{omain}
\eeq
where
\beqn
\mathcal{F} (\bk) &=&  \left( \frac{\partial
\varepsilon_{\bk+\bQ}}{\partial \bk} \times \frac{\partial
\varepsilon_\bk}{\partial \bk} \right) \sum_\omega \frac{2 m^3}{(-i \omega + E_{1\bk})^3 (- i \omega + E_{2\bk})^3} \nn
&=& 6 m^3  \left( \frac{\partial
\varepsilon_{\bk+\bQ}}{\partial \bk} \times \frac{\partial
\varepsilon_\bk}{\partial \bk} \right) \frac{( \mbox{sgn} (E_{1\bk}) - \mbox{sgn}(E_{2 \bk}))}{(E_{1\bk} - E_{2 \bk})^5}.
\label{J3}
\eeqn
In the last step, we have evaluated frequency summation at zero temperature. In the remaining analysis
we will assume we are dealing with a fully gapped insulator with $E_{1\bk} > 0$ and $E_{2 \bk } < 0$
over the entire Brillouin zone. The metallic case has singularities at the Fermi surfaces
which are at $E_{1\bk}=0$ or $E_{2\bk}=0$,
but we will not explore its consequences here; indeed in our expansion in powers of $\bq_{1,2}$, we have implicitly
assumed smooth behavior across the Brillouin zone. Note that in both the insulator and the metal there
is no singularity due to the denomination in Eq.~(\ref{J3}): via Eq.~(\ref{E12}) we always have
$E_{1\bk} - E_{2 \bk} \geq 2 m$.

\begin{figure}[h]
\includegraphics[width=4in]{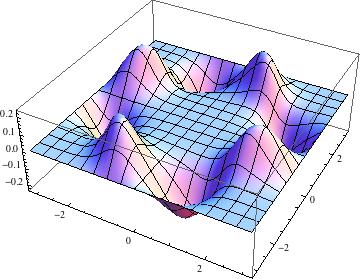}
\caption{A plot of the function $\mathcal{F} (\bk)$ in
Eq.~(\ref{f}) for $\varepsilon_\bk = \cos k_x - \cos k_y + 0.4
\cos (k_x + k_y) + 0.4 \cos(k_x - k_y)$ and $m = 1$.}
\label{fig:f}
\end{figure}

A plot of $\mathcal{F} (\bk)$ for the insulating case is shown in Fig.~\ref{fig:f}.

The integral of $\mathcal{F} (\bk)$ is zero over the Brillouin
zone. However, note that it has the same symmetry as the function
$(\cos k_x - \cos k_y) \sin k_x \sin k_y$; so the integral of
$\mathcal{F} (\bk) (\cos k_x - \cos k_y) \sin k_x \sin k_y$ is
non-zero. This suggest we define the charge $\mathcal{Q}$ by
\beq \mathcal{Q} = - i \sum_{\bk}  c^\dagger (\bk) c ( \bk +
\bQ)  (\cos k_x - \cos k_y) \sin k_x \sin k_y . \eeq Note
$\mathcal{Q}^\dagger = \mathcal{Q}$.

Our main result in Eq.~(\ref{omain}) implies that
any quantum fluctuation which leads to a non-zero value of the
skyrmion density $\epsilon_{abc} n^a \partial_x n^b \partial_y n^c$ will
induce a change in $\mathcal{O}$. Generically, a change in $\mathcal{O}$
must imply a corresponding change in $\mathcal{Q}$ because the two
observables have identical signatures under all symmetries of the Hamiltonian.
In paticular, a hedgehog tunneling event is one in which the spatial
integral of $\epsilon_{abc} n^a \partial_x n^b \partial_y n^c$ (the skyrmion number)
changes by $ 4 \pi$. Thus, before the hedgehog event
$\langle \mathcal{Q} \rangle =0$, while after the hedgehog tunneling
event, we have $\langle \mathcal{Q} \rangle \neq 0$. We can
normalize $\mathcal{Q}$ so that $\langle \mathcal{Q} \rangle
=1$ for each hedgehog, and the normalization constant will depend
upon Eq.~(\ref{J3}) and the details on the band structure. Then with such a
normalization, we have the important correspondence
\beq
\mathcal{Q} \cong \mbox{skyrmion number.}
\label{skyrmion}
\eeq
This is the key result of the present subsection. We emphasize that such
a correspondence is possible because both the skyrmion number
and $\mathcal{Q}$ are invariant under spin rotations, have identical
transformations under all square lattice space group operations, and are both
odd under time-reversal.

\subsection{Connection to VBS order}
\label{sec:cvbs}

The results in
Eqs.~(\ref{omain}) and (\ref{skyrmion})
suggest strong consequences in the `quantum disordered' phase where N\'eel
order has been lost. Such a phase will have a proliferation of
hedgehog tunnelling events, and so Eq.~(\ref{skyrmion})
implies that there will be correspondingly large fluctuations in
the charge $\mathcal{Q}$. We can therefore expect that
fluctuations in variables conjugate to $\mathcal{Q}$ will be
suppressed, and will therefore have long-range order: this is the
competing order induced by the geometric phase in
Eq.~(\ref{main}).
Thus any quantum variable conjugate to $\mathcal{Q}$ is a
bona-fide competing order. There are many possibilities, but
here, we verify that the traditional VBS order does satisfy the
requirements. A more specific field-theoretic discussion of the appearance
of VBS order in the quantum-disordered N\'eel phase will be given in Section~\ref{sec:dual}.

The VBS order is $V=V_x + i V_y$ defined by \beqn
V_x &=& i \sum_{\bk} c^\dagger (\bk) c ( \bk + \bQ_x ) \sin k_x
\nn V_y &=& i \sum_\bk c^\dagger (\bk ) c ( \bk + \bQ_y ) \sin k_y
\label{defvbs} \eeqn where $\bQ_x = (\pi, 0)$ and $\bQ_y = (0,
\pi)$. Now we can compute the commutators \beqn
\left[\mathcal{Q}, V_x \right] &=&  -\sum_\bk c^\dagger (\bk ) c
( \bk + \bQ_y ) \sin k_y \frac{(\cos (k_x) - \cos (3k_x))}{2}
\simeq i V_y \nn \left[ \mathcal{Q},V_y \right] &=&  \sum_\bk
c^\dagger (\bk ) c ( \bk + \bQ_x ) \sin k_x \frac{(\cos (k_y) -
\cos (3k_y))}{2} \simeq -i V_x \eeqn Here the $\simeq$ means that
the two operators have the same symmetry under the square lattice
space group. Thus we have the key result \beq [ \mathcal{Q}, V]
\simeq  V. \label{keyresult}\eeq This means that $V$ is a raising
order for $\mathcal{Q}$. But, as we noted in
Section~\ref{sec:linear}, this is precisely the effect of the
monopole tunneling event: in other words, $V$ has the same quantum
numbers as a monopole operator. Then, following the reasoning in
Refs.~\onlinecite{rs2,senthil2}, we may conclude that $V$ is a
competing order which becomes long-range in the quantum-disordered
N\'eel phase.

An alternative route to determining an operator conjugate to $\mathcal{Q}$
is to determine a $V$ so that $-i (V^\dagger \partial_t V - V \partial_t
V^\dagger) \simeq \mathcal{Q}$. It is easy to check that the
definition in Eq.~(\ref{defvbs}) does satisfy the needed
requirements.
We have the time derivative \beqn \frac{dV_x}{dt}
&=& \sum_\bk \sin k_x (\varepsilon_\bk - \varepsilon_{\bk + \bQ_x}
) c^\dagger (\bk) c (\bk + \bQ_x ) + 2 m \sum_\bk \sin k_x
c^\dagger (\bk) \sigma^z c (\bk + \bQ_y ) \eeqn and similarly for
$V_y$. For simplicitly, we will drop the terms proportional to
$m$, and work in the limit of small $m$. So we have
\beqn
- i
\left( V_y \frac{d V_x}{dt} -   V_x \frac{d V_y}{dt} \right) &=&
\sum_{\bk, \bq} \sin k_x \sin q_y (\varepsilon_\bk -
\varepsilon_{\bk + \bQ_x} ) c^\dagger (\bq) c (\bq + \bQ_y)
c^\dagger (\bk) c (\bk + \bQ_x) \nn &~&~~~~~-  (x \leftrightarrow
y) \label{dtv1}
\eeqn
Now we can factorize the 4-Fermi term using $\langle
c^\dagger (\bk) c (\bk) \rangle = n(\bk)$:
\beqn
- i \left( V_y
\frac{d V_x}{dt} -   V_x \frac{d V_y}{dt} \right) &=& \sum_{\bk}
\sin k_x \sin k_y c^\dagger (\bk ) c (\bk + \bQ) \Bigl[
(\varepsilon_{\bk + \bQ_y} - \varepsilon_{\bk+ \bQ})(1-n(\bk +
\bQ_y))  \nn &~&~~~~~~~~- (\varepsilon_\bk - \varepsilon_{\bk +
\bQ_x}) n( \bk + \bQ_x) \Bigr] - (x \leftrightarrow y) \nn &=&
\sum_{\bk} \sin k_x \sin k_y c^\dagger (\bk ) c (\bk + \bQ) \Bigl[
\varepsilon_{\bk + \bQ_y} - \varepsilon_{\bk+ \bQ_x} \nn &~&~~~~~+
2 (\varepsilon_{\bk + \bQ_x} n (\bk + \bQ_x) - \varepsilon_{\bk +
\bQ_y} n (\bk + \bQ_y)) \nn &~&~~~~~-
 (\varepsilon_\bk + \varepsilon_{\bk + \bQ}) ( n( \bk + \bQ_x) - n (\bk + \bQ_y)) \Bigr]
 \label{dtv2}
\eeqn
The r.h.s. is indeed $\simeq \mathcal{Q}$.

\section{Fluctuating N\'eel and VBS orders}
\label{sec:liang}

Given the connection between the skyrmion number of the N\'eel order and the VBS order
derived in Section~\ref{sec:neel}, it is natural to wonder whether the two order parameters
can be treated at a more equal footing. In Section~\ref{sec:neel} we investigated the fermion correlations
in the background of a spatially varying N\'eel order, and so this suggests a natural
generalization in which we allow for a background spacetime dependence of {\em both} the N\'eel
and VBS orders. This section will present the needed generalization. The result here will be an alternative
derivation of the arguments of Section~\ref{sec:cvbs}: the skyrmion number of the N\'eel order
and the angular variable, $\phi$, of VBS order $V \sim e^{i \phi}$
are quantum-mechanically conjugate variables.

We start from a Neel state with the order parameter $\bm=m (n^x, n^y, n^z) \neq 0$.
When the system approaches the Neel-VBS transition, fluctuating
VBS order becomes important and needs to be taken into account.
The starting point of our analysis is the electron Hamiltonian $H$ with both $\bm$ and $V$:
\begin{eqnarray}
H(V_x, V_y, n^x, n^y) &=& [ H_b + m H^N_z ] + [ V_x H^V_x + V_y H^V_y  + m (n^x H^N_x + n^y H^N_y) ]
\end{eqnarray}
where $H_b$ describes the electron band structure in the absence of Neel or VBS order as in Eq.~(\ref{e1});
the fermion bilinear operator $(H^V_x, H^V_y)$ is dimerized electron hopping
in $x$ and $y$ directions; $(H^N_x, H^N_y, H^N_z)$ is staggered electron spin density in $x,y,z$ directions in
spin space;
$(V_x, V_y)$ describes the fluctuating VBS order; $(n^x, n^y)$ describes the Goldstone mode of the N\'eel order.
We now integrate out the fermions and derive the effective action $S$ for the slowly varying bosonic fields
$\mathcal{A}^\mu (x, y, \tau ) \equiv (V_x, V_y, n^x, n^y)$.

Treating the second term in $H$ as a perturbation,
we find couplings between Neel and VBS order starts at fourth order in a one-loop expansion:
\begin{eqnarray}
S_{1} &=& \sum_{\mu, \nu, \lambda, \delta} \int \prod_{i=1}^3 d \bp_i \; K^{\mu\nu\lambda; \delta}_{\bp_1 \bp_2 \bp_3}
\cdot \mathcal{A}^{\mu}(\bp_1) \mathcal{A}^{\nu}(\bp_2) \mathcal{A}^{\lambda}(\bp_3) \mathcal{A}^{\delta}(-\bp_1 - \bp_2 - \bp_3) ,  \label{s}
 \end{eqnarray}
[Note: we drop all numerical prefactors in this subsection.]
Here $\mu, \nu, \lambda, \delta=1,...4$ labels the components of the perturbation field $\bA$ and the vertex;
$\bp=(p^0, p^x, p^y)$ is the external momenta of $\bA$.
We now expand the function $K$ in powers of $\bp$ and collect terms involving the product $p^\alpha_1 p^\beta_2 p^\gamma_3$ with $\alpha, \beta, \gamma=0,x,y$:
\begin{eqnarray}
K^{\mu\nu\lambda; \delta}_{\bp_1 \bp_2 \bp_3} =K^{\mu\nu\lambda; \delta}_{\alpha\beta\gamma} \cdot p^\alpha_1 p^\beta_2 p^\gamma_3 + ...
\end{eqnarray}
This corresponds to a derivative expansion in real spacetime:
\begin{eqnarray}
S_{1} =  \sum_{\mu, \nu, \lambda, \delta} K^{\mu\nu\lambda; \delta}_{\alpha\beta\gamma} \int d x d y d \tau \;
(\mathcal{A}^{\delta} \partial_\alpha \mathcal{A}^{\mu}  \partial_\beta \mathcal{A}^{\nu} \partial_\gamma \mathcal{A}^{\lambda})  \label{S}
\end{eqnarray}

The action (\ref{S}) resembles the Chern-Simons theory in 6+1 dimensions. A difference is that
the space-time indices $\alpha,\beta,\gamma$ and the internal indices $\mu,\nu,\delta, \lambda$ do not mix with each other.
Qi {\it et al.\/} \cite{qi} recently proposed that  $S_1$ can be simply obtained from the Chern-Simons term by dimensional reduction to 2+1 dimensions. The procedure is to throw away all components in the Chern-Simons term, which involve
spatial derivatives in the internal dimension.
We shall show by calculating $K^{\mu\nu\lambda; \delta}_{\alpha\beta\gamma}$ explicitly that this dimensional reduction approach
does not apply in the present situation.

Among the terms in $S_1$, we are particularly interested in a topological term
\begin{eqnarray}
S_{top} =\sum_{\alpha \beta} i K_{\alpha\beta} \int d x d y d \tau \;  j^N_\alpha j^V_\beta,
\end{eqnarray}
where $j^N_\alpha$ is the skyrmion current in the Neel state:
\begin{eqnarray}
j^N_\alpha \equiv \epsilon_{\alpha \beta \gamma}  \epsilon_{abc} \bn  \partial_\beta n^b \partial_\gamma n^c ,
\label{skyrmioncurrent}
\end{eqnarray}
and $j^V_\beta$ is the VBS current:
\begin{eqnarray}
j^V_\beta \equiv V_x \partial_\beta V_y  - V_y \partial_\beta V_x.
\label{vbscurrent}
\end{eqnarray}

\begin{figure}
\centering
\includegraphics[width=5.5in]{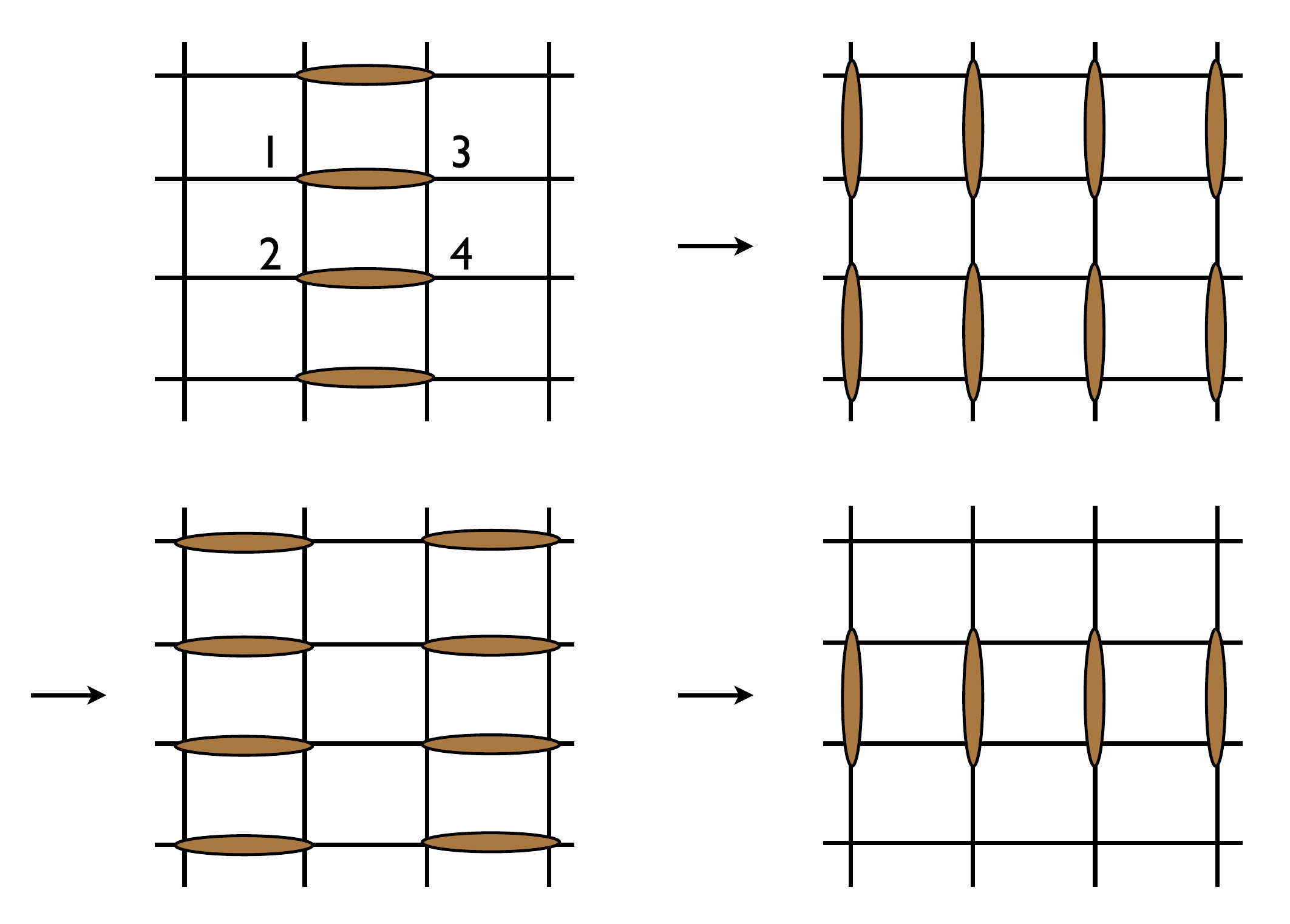}
\caption{VBS order on the  square lattice.}
\label{fig:vbs}
\end{figure}

It follows from symmetry analysis that on the square and honeycomb lattice,
The matrix $K_{\alpha\beta}$ is diagonal. Because of four- and six-fold rotational symmetry, $K_{xx}=K_{yy}$.
 $S_{top}$ then becomes
\begin{eqnarray}
S_{top} = i  \int d x d y d \tau \;  (K   j^N_t j^V_t +  K' j^N_x j^V_x + K' j^N_y j^V_y), \label{jj}
\end{eqnarray}
Comparing (\ref{jj}) and (\ref{S}), we can express $K$ in terms of the tensor components $K^{\mu\nu\lambda; \delta}_{\alpha\beta\gamma}$:
\begin{eqnarray}
K&\propto& [ K^{234;1}_{0xy} +   \textrm{Permutations of (2,0), (3,$x$), and (4,$y$)}  ]  \nonumber \\
&- &  [ K^{243;1}_{0xy}+ \textrm{Permutations of (2,0), (4,$x$), and (3,$y$)} ] \nonumber \\
&-& [ K^{134;2}_{0xy}   + \textrm{Permutations of (1,0), (3,$x$), and (4,$y$)} ]  \nonumber \\
&+& [ K^{143;2}_{0xy} + \textrm{Permutations of (1,0), (4,$x$), and (3,$y$)}  ]
\end{eqnarray}
We now calculate $K$ for the square lattice.  The Hamiltonian $H_b$ is specified in Eq.~(\ref{e1}), and we choose only nearest
neighbor hopping $t$. For the coupling to the order parameters, we choose
\begin{eqnarray}
&& H^N_a = \sum_{i \in A} c^\dagger (\br_i) \sigma^a c (\br_i ) - \sum_{i \in B} c^\dagger (\br_i) \sigma^a c (\br_i )  \nonumber \\
&& H^{V}_\beta =  \sum_{i \in A} (-1)^{i_\beta} [ c^\dagger (\br_i ) c ( \br_i+ {\bf e}_\beta ) + c.c.] , \; \beta=x,y
\end{eqnarray}
where we have divided the square lattice into two sublattices A and B defined by
$(-1)^{i_x+i_y}=\pm 1$.
The N\'eel order carries crystal momentum $(\pi, \pi)$.
The VBS order in the $x$- and $y$-directions carries crystal momentum $(\pi, 0)$ and $(0, \pi)$ respectively,
with the corresponding dimerization pattern shown in Fig.~\ref{fig:vbs}.
Note that the dimerization pattern ``rotates'' around a site as the phase of $V_x + i V_y$ advances by $2\pi$.
It is straightforward to check that the term $S_{top}$ in Eq.(\ref{S}) satisfies  square lattice symmetry.

\begin{figure}
\centering
\includegraphics[width=3.5in]{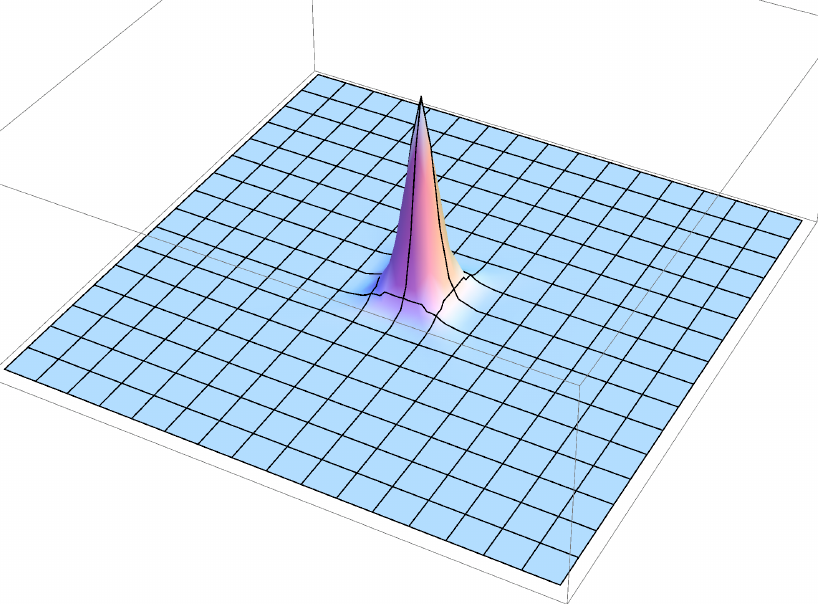}
\caption{Plot of $\cal K$ over a quarter of the  Brillouin zone for $t=1, m=0.25$. }
\label{fig:kplot}
\end{figure}

The lattice periodicity is  doubled in both the $x$ and $y$ direction for $m \neq 0$ and $V_x, V_y \neq 0$.
We choose the 4 sites in a plaquette as the new unit cell.
The Bloch Hamiltonian $H(k_x, k_y)$ is obtained by Fourier transform:
\begin{eqnarray}
H(k_x, k_y )&=& \left(
\begin{array}{cccc}
\bm  \bs & t_y   &  t_x   & 0 \\
t_y^*   & -\bm  \bs  & 0 & t_x   \\
 t_x^*    & 0 & -\bm  \bs & t_y   \\
 0 & t_x^*   & t_y^*   & \bm  \bs
\end{array}
\right) \nonumber \\
t_y &=& 
-t  \cos k_y + i V_y  \sin k_y,  \nonumber \\
t_x &=&  
-t  \cos k_x + i V_x \sin k_x.
\end{eqnarray}
Here $k_x, k_y \in [-\pi/2, \pi/2]$ is crystal momentum in the folded Brilluoin zone.
We used Mathematica to evaluate $K$ and found
\begin{eqnarray}
K=   \int d k_0 d k_x d k_y \frac{m^3 t^2 \sin^2 k_x \sin^2 k_y [ t^4(\cos^2 k_x - \cos^2 k_y)^2-  (k_0^2 + m^2)^2  ]}
{[t^2(\cos k_x - \cos k_y)^2 + k^2_0 + m^2 ]^3 [t^2 (\cos k_x + \cos k_y)^2 + k^2_0 + m^2]^3}.
\end{eqnarray}
The integration over $k_0$ can be done analytically using Mathematica.
The resulting integrand ${\cal K}(k_x, k_y)$ is a complicated function of $k_x$ and $k_y$.
Instead of showing its explicit form, we plot ${\cal K}(k_x, k_y)$ over the Brillouin zone $k_x, k_y \in [0, \pi]$ in Fig.~\ref{fig:kplot}
Note that the integrand is peaked at the ``hot spot'' $Q=(\pi/2, \pi/2)$.
This is not surprising because both the N\'eel and VBS orders have strong nesting at $Q$.

The other coefficient $K'$ in $S_1$ can be obtained similarly and is given by:
\begin{eqnarray}
K'= \int d k_0 d k_x d k_y \frac{m^3 t^2 \sin^2 k_x \sin^2 k_y [ t^4(3\cos^2 k_x + \cos^2 k_y)
(\cos^2 k_x + 3\cos^2 k_y) - 3 (k_0^2 + m^2)^2  ]}
{[t^2(\cos k_x - \cos k_y)^2 + k^2_0 + m^2 ]^3 [t^2 (\cos k_x + \cos k_y)^2 + k^2_0 + m^2]^3}. \nonumber
\end{eqnarray}
Comparing $K$ and $K'$, we found that in general $K \neq K'$. 
This means that different terms in the effective action (\ref{S}) have different coefficients, so that
$S$ cannot be obtained by dimensional reduction from a Chern-Simons term in 6+1 dimensions, which has a single coefficient.

\subsection{Quantum disordered N\'eel phase}
\label{sec:dual}

We can now use the results of this section to
present an alternative version of the argument in Section~\ref{sec:cvbs},
that the quantum disordered N\'eel phase has VBS order. The argument here will be
closer in spirit to the duality mapping discussed in
Ref.~\onlinecite{rs2}.

We will limit our discussion to a quantum-disordered N\'eel phase where the
monopole density is very dilute. Thus, we assume that over a significant intermediate
length scale there is an effective description in terms of a theory in which the total Skyrmion
number is conserved.
As discussed in Section~\ref{sec:neel} and Appendix~\ref{app:gauge}, we can represent
the fluctuations in the local Skyrmion density by a  low energy U(1) photon field $A_\alpha$:
by Eq.~(\ref{gaugeskyrmion}), the gauge flux in this field, $\epsilon_{\alpha\beta\gamma} \partial_\beta A_\gamma$,
is a quarter the Skyrmion current $j^{N}_\alpha$ in Eq.~(\ref{skyrmioncurrent}).
We can write an effective action of the photons as
\beq
\mathcal{L}_{\rm eff}
=  \frac{1}{2 e^2} \left(
\epsilon_{\alpha\beta\gamma} \partial_\beta A_\gamma \right)^2  +
4 i K j^{V}_\alpha \epsilon_{\alpha\beta\gamma} \partial_\beta A_\gamma.
\label{Qmu}
\eeq
Here the second term represents the topological term in Eq.~(\ref{jj}). 
For simplicity, we have assumed $K'=K$. Different values of $K$ and $K'$ will not 
affect our conclusion below. 
Also note that by the discussion at the end of Section~\ref{sec:cvbs},
$\mathcal{Q}$ is conjugate to $j_0^V$.

Now let us perform the standard
duality transformation of 2+1 dimensional electrodynamics
\cite{rs2,senthil2,polyakov} on $\mathcal{L}_{\rm eff}$. The first step
corresponds to decoupling the Maxwell term by a
Hubbard-Stratonovich field, $J_\alpha$, to obtain
\beq
\mathcal{L}_{\rm eff}
=  \frac{e^2}{8 \pi^2} J_\alpha^2 + \frac{i}{2 \pi} J_\alpha
\epsilon_{\alpha\beta\gamma} \partial_\beta A_\gamma   +
4 i K j^{V}_\alpha \epsilon_{\alpha\beta\gamma} \partial_\beta A_\gamma.
\label{Qmu2}
\eeq
Now, we integrate over $A_\alpha$, and this yields the constraint
\begin{equation}
J_\alpha = \partial_\alpha \phi - 8 \pi K \,  j^V_\alpha .
\label{Qmu3}
\end{equation}
where $\phi$ is the scalar field which is dual to the photon. We have judiciously chosen
factors of $(2\pi)$ above to ensure a normalization so that $e^{i \phi}$ is the monopole operator.
Finally, inserting Eq.~(\ref{Qmu3}) into (\ref{Qmu2}) we obtain
\beq \mathcal{L}_{\rm eff} =
\frac{e^2}{8 \pi^2}
\left(\partial_\alpha \phi -  8 \pi K \, j^V_\alpha \right)^2 . \label{lphoton}
\eeq

The effective Lagrangian for the photon phase in Eq.~(\ref{lphoton}) allows to
conclude that the long-range correlations of $\partial_\alpha \phi$ have the same form as those
of $j^V_\alpha$. In other words, we have the
operator correspondence $\partial_\alpha \phi \simeq
j_\alpha^V$. In terms of the complex VBS order parameter $V = V_x + i V_y$
we can therefore write for the monopole operator
$e^{i \phi} \sim V^\nu$, where $\nu$ in general appears to be an irrational number.
In the special cases where the value of $K$ was quantized by projection to an integer number
of electrons per site \cite{haldane,rs1,rs2,senthil1,senthil2}, $\nu$ was found to be an integer;
this is a possible value of $\nu$ here, although our present methods don't allow us to see why any
particular integer would be preferred. The uncertainty in the value of $\nu$ here is analogous to the
arbitrariness in the overall normalization of $\mathcal{Q}$ in Section~\ref{sec:linear}.

In any case, as long as $\nu$ is not an even integer, the correspondence between the
monopole operator $e^{i \phi}$ and $V^\nu$ implies that $V$ has long-range correlations in the monopole-free
region. At even longer scales, once the monpoles condense, the phase of $V$ is locked along one of the lattice
directions \cite{senthil1,senthil2}.

\section{Honeycomb lattice}
\label{sec:honeycomb}

This section will apply the methods developed in
Section~\ref{sec:neel} and Appendix~\ref{app:gauge} to the honeycomb
lattice. As is well known, this lattice has an electronic
dispersion with a Dirac form at low energies. We will adapt our
methods to the Dirac fermions, and find that many results can be
computed rapidly in closed form.

The honeycomb lattice has 2 sublattices, and we label the fermions
on two sublattices as $c_A$ and $c_B$.
To begin, we only include N\'eel order explicitly.
Then the analog of
Eq.~(\ref{e2}) is
\begin{equation}
H = - t\sum_{\langle ij \rangle} \left( c_{Ai}^{\dagger} c_{B j} + c_{B j }^{\dagger} c_{A i}
\right) + m \sum_{i \in A} c_{Ai}^{\dagger} n^a (\br_i ) \sigma^a c_{A i }
- m \sum_{i \in B} c_{Bi}^{\dagger} n^a (\br_i ) \sigma^a c_{B i }.
\end{equation}
We restrict to the case with constant N\'eel order $n^a$,
transform to momentum space, and introduce Pauli matrices
$\tau^a$ in sublattice space, and obtain
\beqn
H &=& \int
\frac{d^2 k}{4 \pi^2} c^{\dagger} ({\bf k}) \Bigl[  -t
\Bigl(\cos({\bf k} \cdot {\bf e}_1) + \cos({\bf k} \cdot {\bf
e}_2)
+ \cos({\bf k} \cdot {\bf e}_3) \Bigr) \tau^x \nonumber \\
&~&~~~~~~+ t \Bigl(\sin({\bf k} \cdot {\bf e}_1) + \sin({\bf k}
\cdot {\bf e}_2) + \sin ({\bf k} \cdot {\bf e}_3) \Bigr) \tau^y +
m \tau^z n^a \sigma^a \Bigr] c ({\bf k} )
\label{hlat}
\eeqn
where we have introduced the unit length vectors \beq \ea = (1,0) \quad
, \quad \eb = (-1/2, \sqrt{3}/2) \quad , \quad \ec = (-1/2,
-\sqrt{3}/2). \eeq We also note that we take the origin of
co-ordinates of the honeycomb lattice at the center of an empty
hexagon, so the A sublattice sites closest to the origin are at
$\ea$, $\eb$, and $\ec$, while the B sublattice sites closet to
the origin are at  $-\ea$, $-\eb$, and $-\ec$.

The low energy electronic excitations reside in the vicinity of
the wavevectors $\pm \bQ_1$, where $\bQ_1 = (4 \pi/9) (\eb -
\ec)$. So we take the continuum limit in terms of the 8-component
field $C$ defined by
\beq
C_{A1} = c_A ({\bf Q}_1) \quad,
\quad C_{B1} = c_B ({\bf Q}_1) \quad , \quad C_{A2} =
c_A (-{\bf Q}_1) \quad, \quad C_{B2} = c_B (- {\bf Q}_1).
\label{cC}
\eeq
In terms of $C$, we obtain from Eq.~(\ref{hlat})
\beq H =
\int \frac{d^2 k}{4 \pi^2} C^{\dagger} ({\bf k} ) \Bigl( v
\tau^y k_x + v \tau^x \rho^z k_y + m \tau^z n^a \sigma^a \Bigr) C
({\bf k}), \label{ham}
\eeq
where $v=3t/2$; below we set $v=1$. We
have also introduced Pauli matrices $\rho^a$ which act in the
$1,2$ valley space. This is the final form of $H$: it makes the
Dirac structure evident, and will also be the most convenient for
our computations.

It is also convenient to list the effects of various symmetry
operations on $C$. Under reflections, $\mathcal{I}_y$, which
sends $x \leftrightarrow -x$ \beq \mathcal{I}_y: \quad C_{A1}
\rightarrow C_{B1} \quad, \quad C_{B1} \rightarrow C_{A1}
\quad , \quad C_{A2} \rightarrow C_{B2} \quad, \quad
C_{B2} \rightarrow C_{A2} \eeq Similarly \beq \mathcal{I}_x:
\quad C_{A1} \rightarrow C_{A2} \quad, \quad C_{B1}
\rightarrow C_{B2} \quad , \quad C_{A2} \rightarrow
C_{A1} \quad, \quad C_{B2} \rightarrow C_{B1} \eeq
Rotations by 60 degrees, $R$, lead to \beq R: \quad C_{A1}
\rightarrow \omega^2 C_{B2} \quad, \quad C_{B1} \rightarrow
\omega C_{A2} \quad , \quad C_{A2} \rightarrow \omega
C_{B1} \quad, \quad C_{B2} \rightarrow \omega^2 C_{A1}
\eeq Translation by the unit cell vector ${\bf e}_2 - {\bf e}_3$,
$T_y$: \beq T_y: \quad C_{A1} \rightarrow \omega^2 C_{A1}
\quad, \quad C_{B1} \rightarrow \omega^2 C_{B1} \quad ,
\quad C_{A2} \rightarrow \omega C_{A2} \quad, \quad
C_{B2} \rightarrow \omega C_{B2} \eeq Time reversal $t
\rightarrow -t$: \beqn \mathcal{T}  : \quad C_{A1} \rightarrow
i\sigma^y C_{A2} \quad  , \quad  C_{A2} \rightarrow
i\sigma^y C_{A1} \quad  , \quad  C_{B1} \rightarrow
i\sigma^y C_{B2} , \quad C_{B2} \rightarrow i\sigma^y
C_{B1} \eeqn Notice that time reversal transformation also
involves a complex conjugation transformation.

\begin{figure}
\centering
\includegraphics[width=4in]{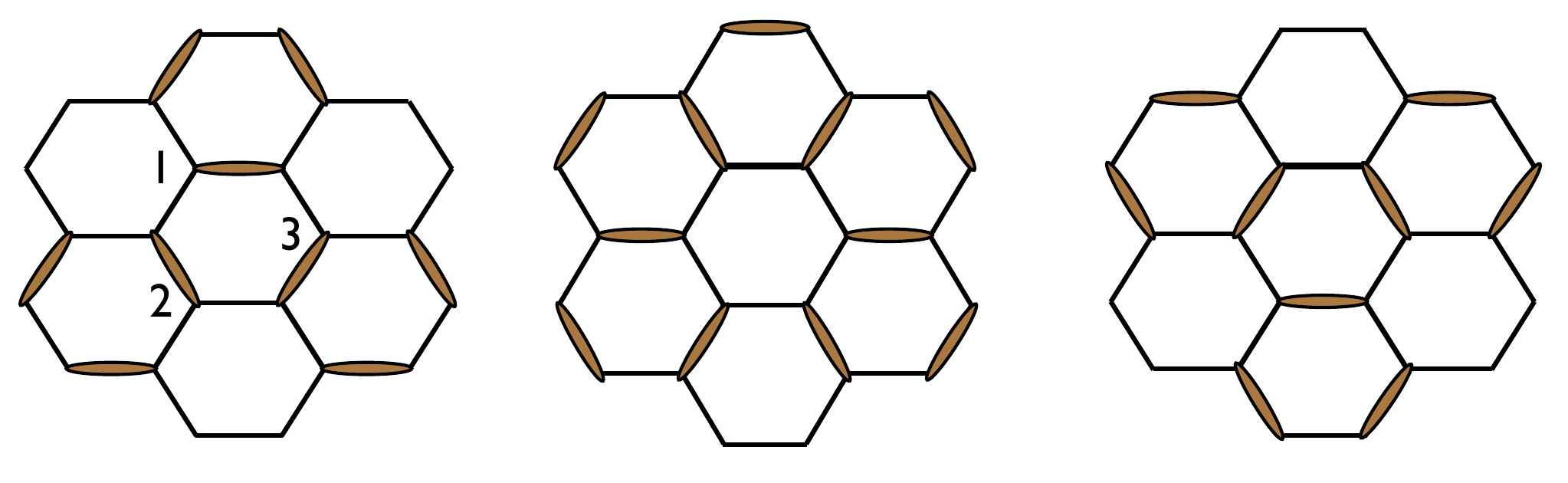}
\caption{VBS order on the honeycomb lattice.}
\label{fig:honeycombvbs}
\end{figure}

From these transformations, we can construct the fermion
bilinear associated with the kekule VBS pattern shown in Fig~\ref{fig:honeycombvbs}.
In terms of the continuum field
$C$, the VBS order parameter is
\beq
V  =
C^\dagger \tau^x (\rho^x + i \rho^y) C  \label{VBShc}
\eeq
We can verify this is the VBS order with the kekule pattern of Fig.~\ref{fig:honeycombvbs}
by its symmetry
transformations
\beqn
\mathcal{I}_y&:& V \rightarrow V \nonumber \\
\mathcal{I}_x&:& V \rightarrow V^* \nonumber \\
R &:& V \rightarrow V^* \nonumber \\
T_y &:& V \rightarrow \omega^2 V \nonumber \\ \mathcal{T} &:& V \rightarrow V^\ast.
\eeqn

\subsection{6D method}
\label{sec:6D}

In the present situation with a Dirac fermion spectrum, the dimensional reduction method \cite{yaolee,qi}
from 6D does apply, and be used to compute the coupling
between the fluctuating N\'eel and VBS orders.
From Eqs. (\ref{ham}) and (\ref{VBShc}), we can write
down the explicit form of the Hamiltonian in the $8 \times 8$ space
of Dirac fermions:
\beqn
&& H (\bk, k_1, k_2, k_3, k_4) = \nn
&&~~~~\tau^y k_x + \tau^x \rho^z k_y + \tau^x \rho^x k_1 + \tau^x \rho^y k_2 +
 \tau^z \sigma^x k_3 +  \tau^z \sigma^y k_4
+ m \tau^z \sigma^z n^z .
\label{hd1}
\eeqn
Here $\bk = (k_x, k_y)$, and the `extra-dimensional' momenta $k_{1,2,3,4}$ are related to the order
parameters: $k_1 = V_x$, $k_2 = V_y$, $k_3 = n^x$, and $k_4 = n^y$.
Now note that the matrices in all the terms in Eq.~(\ref{hd1}) anti-commute with each other.
So this has the natural interpretation as 6D Dirac Hamiltonian, where the last term proportional
to $m$ has the interpretation of a Dirac fermion mass.
We can now proceed as in Ref.~\onlinecite{abanov}, and derive the WZW term for the order parameters.

\subsection{U(1) gauge theory}
\label{sec:u1guage}

Next we turn to the analog of the analysis in Section~\ref{sec:linear} for the square
lattice. However, rather than working with the spatially varying N\'eel order as in
Eq.~(\ref{e2}), we will use the U(1) gauge field formulation of Eq.~(\ref{e3}) which
was applied to the square lattice in Appendix~\ref{app:gauge}.

We begin with the $\psi$ fermion Hamiltonian in Eq.~(\ref{e3}) and take
its continuum limit as in Eq.~(\ref{ham}). For this, we make the analog of the transformation in
Eq.~(\ref{cC}) from the lattice $\psi$ fermions to continuum $\Psi$ fermions.
In this manner, we obtain the continuum U(1) gauge theory
\begin{equation}
\mathcal{L} = \Psi^\dagger \Bigl( \partial_\tau - i \sigma^z
A_\tau - i  \tau^y (\partial_x - i \sigma^z A_x) - i  \tau^x
\rho^z ( \partial_y - i \sigma^z A_y )  + m \tau^z \sigma^z
\Bigr) \Psi
\label{LPsi}
\end{equation}
Now we obtain the result which is the analog of Eq.~(\ref{main})
by applying the Kubo formula to determine the response in an
arbitrary fermion bilinear $\Psi^\dagger \Gamma \Psi$ due to an
arbitrary slowly varying $A_\alpha$. This involves evaluating a
diagram with one fermion loop, and the long wavelength result is
\beqn
\left\langle \Psi^\dagger \Gamma \Psi \right \rangle  &=&
\frac{1}{8 \pi} \Bigl\{ (\partial_x A_\tau - \partial_\tau A_x)
\mbox{Tr} \left[ \Gamma \tau^x \right] + (\partial_\tau A_y -
\partial_y A_\tau ) \mbox{Tr} \left[ \Gamma \rho^z \tau^y \right] \nn
&~&~~~~~~~~~~~~~~~+ (\partial_y A_x - \partial_x A_y)  \mbox{Tr}
\left[ \Gamma \rho^z \right] \Bigr\} . \label{kubo}
\eeqn
Now we see that the choices $\Gamma = \tau^x$, $\rho^z \tau^y$, and $\rho^z$
lead to non-zero fermion bilinears induced by the $A_\alpha$ gauge flux.
Note that this result was obtained
with much greater ease in the continuum Dirac theory than for Eq.~(\ref{main}).

Let us restate the result in Eq.~(\ref{kubo}) in different terms. We add to $\mathcal{L}$
in Eq.~(\ref{LPsi}) a source term $j^V_\alpha$:
\beqn
\mathcal{L} \rightarrow \mathcal{L} - \frac{i}{2} \Bigl( j^V_0 \Psi^\dagger
\rho^z \Psi +j^V_x \Psi^\dagger \rho^z
\tau^y \Psi + j^V_y \Psi^\dagger \tau^x \Psi \Bigr).
\label{kubo2}
\eeqn
Then the implication of Eq.~(\ref{kubo2}) is that if we integrate out the $\Psi$ fermions,
the effective action for $j^V_\alpha$ and the gauge field $A_\alpha$ has a mutual Chern-Simons term:
\beqn
\mathcal{L}_{\rm eff} = \frac{1}{12 \pi m} (\epsilon_{\alpha\beta\gamma} \partial_\beta A_\gamma )^2 +
\frac{i}{2 \pi} j^V_\alpha \epsilon_{\alpha\beta\gamma} \partial_\beta A_\gamma
\eeqn
The similarity to Eq.~(\ref{Qmu2}) should now be evident. We can now proceed with the duality
of electrodynamics to obtain the analog of Eq.~(\ref{lphoton}), which is
\beqn
\mathcal{L}_{\rm eff} = \frac{3 m}{8 \pi} \left( \partial_\alpha \phi - j^V_\alpha \right)^2,
\eeqn
where again $e^{i \phi}$ is the monopole operator. As argued below Eq.~(\ref{lphoton}) any
operator with the same quantum numbers as $e^{i \phi}$ has long-range order
in the `quantum-disordered' phase.

Here we present a different route to identifying candidates for the competing order.
First, we notice that the theory in Eq.~(\ref{kubo2}) actually enjoys a gauge invariance under which
\beqn
\Psi \rightarrow \exp \left( i \frac{ \rho^z}{2} \theta \right) \Psi
\quad , \quad j^V_\alpha \rightarrow j^v_\alpha - \partial_\alpha \theta \label{g2}
\eeqn
where $\theta$ is a field with an arbitrary spacetime dependence. (Note that this gauge invariance
is completely different from that associated with the $A_\alpha$ gauge field, under which $\Psi
\rightarrow \exp( (i/2) \sigma^z \theta' ) \Psi$.) Now we observe that this gauge invariance extends
also to Eq.~(\ref{g2}), under which
\beqn
e^{i\phi} \rightarrow e^{i \theta} e^{i \phi}. \label{g3}
\eeqn
We will use Eq.~(\ref{g3}) as the key relation needed for any competing order associated
with the monopole operator $e^{i \phi}$.

Equivalently, we can use Eq.~(\ref{g2}), and
restate the requirement of Eq.~(\ref{g3}) as the commutation relation
\begin{equation}
[\mathcal{Q}, e^{i \phi (\bx)} ] =  e^{i \phi (\bx)}, \label{QV}
\end{equation}
where \beqn \mathcal{Q} = \frac{1}{2} \int d^2 r \, \Psi^\dagger
\rho^z \Psi . \label{fluxquantum} \eeqn This makes a very explicit
connection to Section~\ref{sec:cvbs} and Eq.~(\ref{keyresult}).
Note that here the overall normalization of $\mathcal{Q}$ is
specified, and does not suffer from the arbitrariness we
encountered in Sections~\ref{sec:linear} and~\ref{sec:dual}.

Now we can easily check that the VBS order parameter in Eq.~(\ref{VBShc}) obeys
 the commutation relation
\begin{equation}
[\mathcal{Q}, V({\bf x}) ] =  V({\bf x}), \label{QV2}
\end{equation}
and so we conclude that $e^{i \phi} \simeq V$, and that VBS order can appear in the
quantum-disordered N\'eel phase.

\subsection{Other competing orders}

In addition to the VBS order parameter $V$, it is now easy to see that
there are other order parameters which are canonically conjugate to $\mathcal{Q}$.
For instance,
the following three complex order parameters all satisfy
Eq.~\ref{QV2}:
\beqn
V_1 \sim \Psi^\dagger (\rho^x + i\rho^y) \Psi,
\ V_2 \sim \Psi^\dagger \tau^z (\rho^x + i\rho^y) \Psi, \ V_3 \sim
\Psi^\dagger \tau^y (\rho^x + i\rho^y) \Psi ).
\eeqn
Under
discrete symmetries, these order parameters transform as
\beqn
\mathcal{I}_y&:& V_1 \rightarrow V_1 \quad ,
\quad V_2 \rightarrow - V_2 \quad , \quad V_3 \rightarrow - V_3, \\
\mathcal{I}_x&:& V_\mu \rightarrow V^\ast_\mu \quad \\
R &:& \mathrm{Re}[V_1] + i \mathrm{Im}[V_2] \rightarrow \omega^2
(\mathrm{Re}[V_1] + i \mathrm{Im}[V_2]), \\ && \mathrm{Re}[V_2] +
i \mathrm{Im}[V_1] \rightarrow - \omega^2
 (\mathrm{Re}[V_2] + i
\mathrm{Im}[V_1]),\\ && V_3
\rightarrow - V_3^\ast \\
T_y &:& V_\mu  \rightarrow \omega^2 V_\mu \\ T &:& V_1 \rightarrow
V^\ast_1 \quad , \quad V_2 \rightarrow V^\ast_2 \quad , \quad V_3
\rightarrow - V^\ast_3.
\eeqn
According to these transformation
laws, we can identify that $V_1$ is a charge density wave (CDW)
with wave vector $2\mathbf{Q}_1$, $V_2$ is the $A-B$ sublattice
staggered CDW, and $V_3$ is a charge current density wave.

However, notice that the matrices in $V$ in Eq.~(\ref{VBShc})
anticommute with all the matrices in $H$ in Eq.~(\ref{ham});
therefore the VBS state has the lowest fermionic mean field
energy, because the fermion $\Psi$ will acquire a Dirac mass gap
$m \sim \sqrt{m^2 + |V|^2}$. Compared with the VBS order parameter
$V$, the other three order parameters $V_\mu$ have higher mean
field fermion energy, hence are less favorable in energy.

\subsection{Superconductor order parameters}

In addition to the VBS order parameter, the N\'{e}el order can
also have strong competition with superconductor, as long as the
SC order parameters satisfy Eq.~\ref{QV}. In this section we will
focus on spin singlet pairings. Using the quantum number
$\mathcal{Q}$ in Eq.~\ref{fluxquantum} and criterion Eq.~\ref{QV},
it is straightforward to show that the following six groups of SC
order parameters are candidate competing orders of the N\'{e}el
order: \beqn \mathrm{Group \ 1:} && (\Delta_1, \ \Delta_2) \sim
(\mathrm{Re}[\Psi^t i\sigma^y \Psi], \ \mathrm{Im}[\Psi^t
i\sigma^y \rho^z \Psi]), \cr\cr && \Delta_1 \sim \sum_k C_{A,
\mathbf{Q}_1 + k} i\sigma^y C_{A, \mathbf{Q}_1 - k} + C_{B,
\mathbf{Q}_1 + k} i\sigma^y C_{B, \mathbf{Q}_1 - k} \cr\cr && +
C_{A, - \mathbf{Q}_1 + k} i\sigma^y C_{A, - \mathbf{Q}_1 - k} +
C_{B, - \mathbf{Q}_1 + k} i\sigma^y C_{B, - \mathbf{Q}_1 - k} +
H.c. \cr\cr && \Delta_2 \sim \sum_k i C_{A, \mathbf{Q}_1 + k}
i\sigma^y C_{A, \mathbf{Q}_1 - k} + i C_{B, \mathbf{Q}_1 + k}
i\sigma^y C_{B, \mathbf{Q}_1 - k} \cr\cr && - i C_{A, -
\mathbf{Q}_1 + k} i\sigma^y C_{A, - \mathbf{Q}_1 - k} - i C_{B, -
\mathbf{Q}_1 + k} i\sigma^y C_{B, - \mathbf{Q}_1 - k} + H.c.
\cr\cr\cr \mathrm{Group \ 2:} && (\Delta_1, \ \Delta_2) \sim
(\mathrm{Im}[\Psi^t i\sigma^y \Psi], \ \mathrm{Re}[\Psi^t
i\sigma^y \rho^z \Psi]), \cr\cr && \Delta_1 \sim \sum_k i C_{A,
\mathbf{Q}_1 + k} i\sigma^y C_{A, \mathbf{Q}_1 - k} + i C_{B,
\mathbf{Q}_1 + k} i\sigma^y C_{B, \mathbf{Q}_1 - k} \cr\cr && + i
C_{A, - \mathbf{Q}_1 + k} i\sigma^y C_{A, - \mathbf{Q}_1 - k} + i
C_{B, - \mathbf{Q}_1 + k} i\sigma^y C_{B, - \mathbf{Q}_1 - k} +
H.c. \cr\cr && \Delta_2 \sim \sum_k C_{A, \mathbf{Q}_1 + k}
i\sigma^y C_{A, \mathbf{Q}_1 - k} + C_{B, \mathbf{Q}_1 + k}
i\sigma^y C_{B, \mathbf{Q}_1 - k} \cr\cr && - C_{A, - \mathbf{Q}_1
+ k} i\sigma^y C_{A, - \mathbf{Q}_1 - k} - C_{B, - \mathbf{Q}_1 +
k} i\sigma^y C_{B, - \mathbf{Q}_1 - k} + H.c.
\cr\cr\cr \mathrm{Group \ 3:} && (\Delta_1, \ \Delta_2) \sim
(\mathrm{Re}[\Psi^t \tau^z i\sigma^y \Psi], \ \mathrm{Im}[\Psi^t
\tau^z i\sigma^y \rho^z \Psi]) \cr\cr && \Delta_1 \sim \sum_k
C_{A, \mathbf{Q}_1 + k} i\sigma^y C_{A, \mathbf{Q}_1 - k} - C_{B,
\mathbf{Q}_1 + k} i\sigma^y C_{B, \mathbf{Q}_1 - k} \cr\cr && +
C_{A, - \mathbf{Q}_1 + k} i\sigma^y C_{A, - \mathbf{Q}_1 - k} -
C_{B, - \mathbf{Q}_1 + k} i\sigma^y C_{B, - \mathbf{Q}_1 - k} +
H.c. \cr\cr && \Delta_2 \sim \sum_k i C_{A, \mathbf{Q}_1 + k}
i\sigma^y C_{A, \mathbf{Q}_1 - k} - i C_{B, \mathbf{Q}_1 + k}
i\sigma^y C_{B, \mathbf{Q}_1 - k} \cr\cr && - i C_{A, -
\mathbf{Q}_1 + k} i\sigma^y C_{A, - \mathbf{Q}_1 - k} + i C_{B, -
\mathbf{Q}_1 + k} i\sigma^y C_{B, - \mathbf{Q}_1 - k} + H.c.
\cr\cr\cr \mathrm{Group \ 4:} && (\Delta_1, \ \Delta_2) \sim
(\mathrm{Im}[\Psi^t \tau^z i\sigma^y \Psi], \ \mathrm{Re}[\Psi^t
\tau^z i\sigma^y \rho^z \Psi]) \cr\cr && \Delta_1 \sim \sum_k i
C_{A, \mathbf{Q}_1 + k} i\sigma^y C_{A, \mathbf{Q}_1 - k} - i
C_{B, \mathbf{Q}_1 + k} i\sigma^y C_{B, \mathbf{Q}_1 - k} \cr\cr
&& + i C_{A, - \mathbf{Q}_1 + k} i\sigma^y C_{A, - \mathbf{Q}_1 -
k} - i C_{B, - \mathbf{Q}_1 + k} i\sigma^y C_{B, - \mathbf{Q}_1 -
k} + H.c. \cr\cr && \Delta_2 \sim \sum_k C_{A, \mathbf{Q}_1 + k}
i\sigma^y C_{A, \mathbf{Q}_1 - k} - C_{B, \mathbf{Q}_1 + k}
i\sigma^y C_{B, \mathbf{Q}_1 - k} \cr\cr && - C_{A, - \mathbf{Q}_1
+ k} i\sigma^y C_{A, - \mathbf{Q}_1 - k} + C_{B, - \mathbf{Q}_1 +
k} i\sigma^y C_{B, - \mathbf{Q}_1 - k} + H.c.
\cr\cr\cr \mathrm{Group \ 5:} && (\Delta_1, \ \Delta_2) \sim
(\mathrm{Re}[\Psi^t \tau^x i\sigma^y \Psi], \ \mathrm{Im}[\Psi^t
\tau^x i\sigma^y \rho^z \Psi]) \cr\cr && \Delta_1 \sim \sum_k
C_{A, \mathbf{Q}_1 + k} i\sigma^y C_{B, \mathbf{Q}_1 - k} + C_{B,
\mathbf{Q}_1 + k} i\sigma^y C_{A, \mathbf{Q}_1 - k} \cr\cr && +
C_{A, - \mathbf{Q}_1 + k} i\sigma^y C_{B, - \mathbf{Q}_1 - k} +
C_{B, - \mathbf{Q}_1 + k} i\sigma^y C_{A, - \mathbf{Q}_1 - k} +
H.c. \cr\cr && \Delta_2 \sim \sum_k i C_{A, \mathbf{Q}_1 + k}
i\sigma^y C_{B, \mathbf{Q}_1 - k} + i C_{B, \mathbf{Q}_1 + k}
i\sigma^y C_{A, \mathbf{Q}_1 - k} \cr\cr && - i C_{A, -
\mathbf{Q}_1 + k} i\sigma^y C_{B, - \mathbf{Q}_1 - k} - i C_{B, -
\mathbf{Q}_1 + k} i\sigma^y C_{A, - \mathbf{Q}_1 - k} + H.c.
\cr\cr\cr \mathrm{Group \ 6:} && (\Delta_1, \ \Delta_2) \sim
(\mathrm{Im}[\Psi^t \tau^x i\sigma^y \Psi], \ \mathrm{Re}[\Psi^t
\tau^x i\sigma^y \rho^z \Psi]) \cr\cr && \Delta_1 \sim \sum_k i
C_{A, \mathbf{Q}_1 + k} i\sigma^y C_{B, \mathbf{Q}_1 - k} + i
C_{B, \mathbf{Q}_1 + k} i\sigma^y C_{A, \mathbf{Q}_1 - k} \cr\cr
&& + i C_{A, - \mathbf{Q}_1 + k} i\sigma^y C_{B, - \mathbf{Q}_1 -
k} + i C_{B, - \mathbf{Q}_1 + k} i\sigma^y C_{A, - \mathbf{Q}_1 -
k} + H.c. \cr\cr && \Delta_2 \sim \sum_k C_{A, \mathbf{Q}_1 + k}
i\sigma^y C_{B, \mathbf{Q}_1 - k} + C_{B, \mathbf{Q}_1 + k}
i\sigma^y C_{A, \mathbf{Q}_1 - k} \cr\cr && - C_{A, - \mathbf{Q}_1
+ k} i\sigma^y C_{B, - \mathbf{Q}_1 - k} - C_{B, - \mathbf{Q}_1 +
k} i\sigma^y C_{A, - \mathbf{Q}_1 - k} + H.c. \eeqn All of these
SC order parameters carry nonzero lattice momentum
$2\mathbf{Q}_1$, and none of them gaps out the Dirac points.
Nevertheless, these SC orders are most likely to be adjacent to
the N\'{e}el order on the phase diagram.

\section{Nambu quasi-particles of $d-$wave superconductor}
\label{sec:nodal}

In this section we will apply the above methods to analyze the
$d-$wave superconductor and its descendants. As in previous
section we will examine the nature of the ``quantum disordered''
phase after loss of antiferromagnetic order. However, we will not
consider the case of commensurate antiferromagnetic ordering at
wavevector ${\bf Q} = (\pi, \pi)$, because it requires
computations we have not explored here. Rather, we will limit
ourselves to the technically easier case of nested spin density
wave order, with a wavevector precisely equal to the separation
between two of the nodal points of the fermionic excitations of
the $d$-wave superconductor. The non-nested case is of
experimental importance, but we will not consider it here.

The nodal
quasi-particles of the $d$-wave superconductor are described by
the following Dirac fermion Lagrangian: \beqn L_{\Psi} &=&
\Psi^\dagger_1 (\partial_\tau - i\frac{v_F}{\sqrt{2}} (\partial_x
+ \partial_y) \tau^z - i \frac{v_\Delta}{\sqrt{2}} (-
\partial_x +
\partial_y) \tau^x) \Psi_1 \cr\cr &+& \Psi^\dagger_2
(\partial_\tau - i\frac{v_F}{\sqrt{2}} (- \partial_x + \partial_y)
\tau^z - i \frac{v_\Delta}{\sqrt{2}} (\partial_x + \partial_y)
\tau^x) \Psi_2. \eeqn $\Psi_1 = (f_1, i\sigma^y f^\dagger_3)^t$,
$\Psi_2 = (f_2, i\sigma^y f^\dagger_4)^t$. $f_1$, $f_2$, $f_3$ and
$f_4$ are quasiparticles at nodal points $(Q, Q)$, $(-Q, Q)$,
$(-Q,-Q)$ and $(Q, -Q)$ respectively. Notice that $Q$ is in
general incommensurate, $v_F$ and $v_\Delta$ are different from
each other.

We assume the system has the symmetry of the square lattice. Under
square lattice discrete symmetry group, the quasi-particle
$\Psi_1$ and $\Psi_2$ transform as: \beqn T_x , && \ \ x
\rightarrow x + 1, \ \ \Psi_1 \rightarrow e^{i Q}\Psi_1, \ \Psi_2
\rightarrow e^{-i Q}\Psi_2; \cr\cr T_y , && \ \ y \rightarrow y +
1, \ \ \Psi_1 \rightarrow e^{i Q}\Psi_1, \ \Psi_2 \rightarrow e^{i
Q}\Psi_2; \cr\cr \mathcal{I}_y, && \ \ x \rightarrow - x, \ \
\Psi_1 \rightarrow \Psi_2, \ \Psi_2 \rightarrow \Psi_1; \cr\cr
\mathcal{I}_x, && \ \ y \rightarrow - y, \ \ \Psi_1 \rightarrow
\sigma^y \tau^y \Psi_2^\dagger, \ \Psi_2 \rightarrow
\sigma^y\tau^y \Psi_1^\dagger; \cr\cr \mathcal{I}_{x - y}, && \ \
x \rightarrow y, \ \  y \rightarrow x, \ \ \Psi_1 \rightarrow i
\tau^z \Psi_1, \ \Psi_2 \rightarrow \sigma^y \tau^x
\Psi^\dagger_2; \cr\cr \mathcal{I}_{x + y}, && \ \ x \rightarrow -
y, \ \  y \rightarrow - x, \ \ \Psi_1 \rightarrow \sigma^y \tau^x
\Psi_1^\dagger, \ \Psi_2 \rightarrow i\tau^z \Psi_2; \cr\cr T, &&
\ \ t \rightarrow - t, \ \ \Psi_1 \rightarrow i\tau^y
\Psi_1^\dagger, \ \Psi_2 \rightarrow i\tau^y \Psi_2^\dagger .
\label{nodalsymmetry} \eeqn Notice that transformations
$\mathcal{I}_{x - y}$ and $\mathcal{I}_{x + y}$ are combined with
a U(1) transformation on the superconductor order parameter:
$\Delta e^{i\theta} \rightarrow \Delta e^{i\theta + i\pi}$.

Now let us consider the spin density wave states that gap out the
nodal quasi-particles $i.e.$ the SDW with wave vector $(2Q, 2Q)$,
and $(2Q, -2Q)$, which can be written as $i\Psi^t_1
\tau^y\sigma^y\vec{\sigma} \Psi_1$ and $i\Psi^t_2
\tau^y\sigma^y\vec{\sigma} \Psi_2$ respectively. In contrast, the
SDW at $(2Q, 0)$ and $(0, 2Q)$ will not gap out the nodes, and
they will be ignored hereafter. It is convenient to introduce the
Majorana fermion $\chi_a$ as $\Psi = \chi_A + i\chi_B$, and there
are in total four different choices of SDW that can gap out the
nodal points: \beqn \vec{\Phi}_1 &=& \mathrm{Re}[i\Psi^t
\tau^y\sigma^y\vec{\sigma} \Psi] \sim (\chi^t i \tau^y\sigma^y
\sigma^x \rho^z \chi, \ \chi^t \tau^y\sigma^y \sigma^y \rho^x
\chi, \ \chi^t i \tau^y\sigma^y \sigma^z \rho^z \chi), \cr\cr
\vec{\Phi}_2 &=& \mathrm{Re}[i\Psi^t \tau^y\sigma^y\vec{\sigma}
\mu^z \Psi] \sim (\chi^t i \tau^y\sigma^y \sigma^x \rho^z \mu^z
\chi, \ \chi^t \tau^y\sigma^y \sigma^y \rho^x \mu^z \chi, \ \chi^t
i \tau^y\sigma^y \sigma^z \rho^z \mu^z \chi), \cr\cr \vec{\Phi}_3
&=& \mathrm{Im}[i\Psi^t \tau^y\sigma^y\vec{\sigma} \Psi] \sim
(\chi^t i \tau^y\sigma^y \sigma^x \rho^x \chi, \ \chi^t
\tau^y\sigma^y \sigma^y \rho^z \chi, \ \chi^t i \tau^y\sigma^y
\sigma^z \rho^x \chi), \cr\cr \vec{\Phi}_4 &=& \mathrm{Im}[i\Psi^t
\tau^y\sigma^y\vec{\sigma} \mu^z \Psi] \sim (\chi^t i
\tau^y\sigma^y \sigma^x \rho^x \mu^z \chi, \ \chi^t \tau^y\sigma^y
\sigma^y \rho^z \mu^z \chi, \ \chi^t i \tau^y\sigma^y \sigma^z
\rho^x\mu^z \chi). \label{nodalsdw}\eeqn The Pauli matrices
$\mu^a$ mix $\Psi_1$ and $\Psi_2$, while the Pauli matrices
$\rho^a$ mix $\chi_A$ and $\chi_B$. In the Majorana Fermion basis,
the SU(2) spin operators are represented by the total
antisymmetric matrices \beqn \vec{S} = (\sigma^x \rho^y, \
\sigma^y, \ \sigma^z\rho^y). \eeqn We can check that all four
vectors $\vec{\Phi}_a$ ($a = 1\cdots 4$) transform as vectors
under $\vec{S}$.

Mow we hope to consider the slowly varying SDW by introducing the
SU(2) gauge field $\sum_{l} A^l_\mu S^l$, which will be Higgsed to
U(1) gauge field with a background nonzero expectation value of
$\Phi_a$: \beqn L_{\chi} &=& \chi^t_1 ((\partial_\tau - i
A^l_0S^l) - i v_F(\partial_X - i A^l_X S^l) \tau^z - i v_\Delta
(\partial_Y - i A^l_Y S^l) \tau^x) \chi_1 \cr\cr &+&
\chi^\dagger_2 ((\partial_\tau - i A^l_0 S^l) - i v_F (\partial_Y
- i A^l_Y S^l) \tau^z - i v_\Delta (\partial_X - i A^l_X S^l)
\tau^x) \chi_2 \cr\cr &+& \Phi^l_a \chi^t T^l_a \chi. \eeqn Now we
have redefined the coordinate $(x + y)/\sqrt{2} \rightarrow X$,
$(- x + y)/\sqrt{2} \rightarrow Y$. The order parameter $\Phi_a^l$
has Higgsed the SU(2) gauge field down to U(1) gauge field
$A^l_\mu$. The matrix $T^l_a$ can be found in Eq.~\ref{nodalsdw}.

Now the flux quantum number can be calculated using the same
techniques developed in the previous sections. We summarize our
results in the following: \beqn \mathrm{Group \ 1} &:& \langle
\Phi^l_{1} \rangle \neq 0, \ \ \mathrm{gauge \ flux \ carries} \
\mathcal{Q} \sim  \chi^t \sigma^y \mu^z \rho^x \chi; \cr\cr
\mathrm{Group \ 2} &:& \langle \Phi^l_{2} \rangle \neq 0, \ \
\mathrm{gauge \ flux \ carries} \ \mathcal{Q} \sim  \ \chi^t
\sigma^y \rho^x \chi, \cr\cr \mathrm{Group \ 3} &:& \langle
\Phi^l_{3} \rangle \neq 0, \ \ \mathrm{gauge \ flux \ carries} \
\mathcal{Q} \sim \ \chi^t \sigma^y \mu^z \rho^z \chi, \cr\cr
\mathrm{Group \ 4} &:& \langle \Phi^l_{4} \rangle \neq 0, \ \
\mathrm{gauge \ flux \ carries} \ \mathcal{Q} \sim  \ \chi^t
\sigma^y \rho^z \chi. \eeqn The quantum number $\mathcal{Q}$
carried by the flux is obviously SU(2) gauge invariant.

The flux condensate will again lead to orders that break certain
lattice symmetry in Eq.~\ref{nodalsymmetry}. The condensate order
parameter $V$ has to satisfy Eq.~\ref{keyresult}. Within all these
order parameters that satisfy Eq.~\ref{keyresult}, we choose the
order parameters that have the lowest nodal quasi-particle mean
field energy, $i.e.$ the order parameters that anticommute with
$T^l_a$. We list our results in the following equation, and for
each group of SDW in Eq.~\ref{nodalsdw} we introduce a five
component vector $\Xi^i_{(a)}$ with $\vec{\Phi}_a \sim
(\Xi^1_{(a)}, \Xi^2_{(a)}, \Xi^3_{(a)})$, $V_a \sim \Xi^4_{(a)} +
i\Xi^5_{(a)}$: \beqn \mathrm{Group \ 1} &:& \Xi_{(1)}^{i = 1, 2,
3} = \vec{\Phi}_1 , \cr \cr && \Xi_{(1)}^{4} =
\frac{1}{\sqrt{2}}\chi^t (\tau^z - \tau^x) \mu^y \chi \sim
\Psi^\dagger (\tau^z - \tau^x) \mu^y \Psi , \cr\cr && \Xi_{(1)}^5
= \chi^t (\tau^z - \tau^x) \mu^x \sigma^y \rho^x \chi \sim
\mathrm{Im} [\Psi^t (\tau^z - \tau^x) \mu^x \sigma^y \Psi ];
\cr\cr\cr \mathrm{Group \ 2} &:& \Xi_{(2)}^{i = 1, 2, 3} =
\vec{\Phi}_2 , \cr \cr && \Xi_{(2)}^4 = \frac{1}{\sqrt{2}}\chi^t
(\tau^z - \tau^x) \rho^y \mu^x \chi \sim \Psi^\dagger(\tau^z -
\tau^x) \mu^x \Psi, \cr \cr && \Xi_{(2)}^5 =
\frac{1}{\sqrt{2}}\chi^t (\tau^z - \tau^x) \mu^x \sigma^y \rho^z
\chi \sim \mathrm{Re} [\Psi^t (\tau^z - \tau^x)\sigma^y\mu^x\Psi];
\cr\cr\cr \mathrm{Group \ 3} &:& \Xi_{(3)}^{i = 1,2,3} =
\vec{\Phi}_3 , \cr \cr && \Xi_{(3)}^4 = \Xi_{(1)}^4 =
\frac{1}{\sqrt{2}}\chi^t (\tau^z - \tau^x) \mu^y \chi \sim
\Psi^\dagger (\tau^z - \tau^x)\mu^y \Psi, \cr \cr && \Xi_{(3)}^5 =
\Xi_{(2)}^5 \frac{1}{\sqrt{2}}\chi^t (\tau^z - \tau^x) \mu^x
\sigma^y \rho^z \chi \sim \mathrm{Re} [\Psi^t (\tau^z - \tau^x)
\mu^x \sigma^y \Psi]; \cr\cr \cr \mathrm{Group \ 4}  &:&
\Xi_{(4)}^{i = 1,2,3} = \vec{\Phi}_4 , \cr \cr && \Xi_{(4)}^4 =
\Xi_{(2)}^4 =  \frac{1}{\sqrt{2}}\chi^t (\tau^z - \tau^x) \rho^y
\mu^x \chi \sim \Psi^\dagger ( \tau^z - \tau^x) \mu^x \Psi. \cr\cr
&& \Xi_{(4)}^5 = \Xi_{(1)}^5 = \frac{1}{\sqrt{2}}\chi^t (\tau^z -
\tau^x) \mu^x \sigma^y \rho^x \chi \sim \mathrm{Im} [\Psi^t
(\tau^z - \tau^x) \mu^x \sigma^y \Psi].  \eeqn With the formalism
developed in Ref.~\cite{abanov}, we can also show that there is a
O(5) WZW term for each group of O(5) vector $\Xi^i_{(a)}$. Both
the WZW term and the gauge flux calculations imply that the SDW
$\Phi^i_{(a)}$ and order parameters $V_a \sim \Xi^4_{(a)} +
i\Xi^5_{(a)}$ are competing with each other, and after suppressing
the SDW $\Phi^i_{(a)}$, the system enters the order with nonzero
$\langle V_a \rangle$ directly.

Now we want to identify the physical meanings of $\Xi_{(a)}^4$ and
$\Xi_{(b)}^5$. Clearly, $\Xi_{(a)}^4$ and $\Xi_{(b)}^5$ are both
density waves of physical quantities, with wave vectors $(2Q, \
0)$ and $(0, \ 2Q)$ respectively. Under lattice symmetry
Eq.~\ref{nodalsymmetry}, $\Xi^4_{(a)}$ and $\Xi^5_{(a)}$
transforms as: \beqn \mathcal{I}_y &:& \Xi_{(1)}^4 \rightarrow -
\Xi_{(1)}^4, \ \ \ \Xi_{(1)}^5 \rightarrow \Xi_{(1)}^5, \ \ \ \
\Xi_{(2)}^4 \rightarrow \Xi_{(2)}^4, \ \ \ \Xi_{(2)}^5 \rightarrow
\Xi_{(2)}^5, \cr\cr \mathcal{I}_x &:& \Xi_{(1)}^4 \rightarrow
\Xi_{(1)}^4, \ \ \ \Xi_{(1)}^5 \rightarrow \Xi_{(1)}^5, \ \ \ \
\Xi_{(2)}^4 \rightarrow \Xi_{(2)}^4, \ \ \ \Xi_{(2)}^5 \rightarrow
- \Xi_{(2)}^5, \cr\cr \mathcal{I}_{x-y} &:& \Xi_{(1)}^4
\leftrightarrow \Xi_{(2)}^5, \ \ \ \Xi_{(1)}^5 \leftrightarrow
\Xi_{(2)}^4, \cr\cr \mathcal{I}_{x + y} &:& \Xi_{(1)}^4
\leftrightarrow \Xi_{(2)}^5, \ \ \ \Xi_{(1)}^5 \leftrightarrow
\Xi_{(2)}^4, \cr\cr T &:& \Xi_{(a)}^i \rightarrow \Xi_{(a)}^i, \ \
i = 4,\ 5. \eeqn According to these transformations, we can make
the following identifications: \beqn && \Xi_{(2)}^4 + i
\Xi_{(1)}^4 = \Xi_{(4)}^4 + i \Xi_{(3)}^4 = \mathrm{VBS \ or \ CDW
\ with \ wave \ vector} \ (2Q, \ 0); \cr\cr && \Xi_{(1)}^5 + i
\Xi_{(2)}^5 = \Xi_{(4)}^5 + i \Xi_{(3)}^5 = \mathrm{VBS \ or \ CDW
\ with \ wave \ vector} \ (0, \ 2Q). \eeqn These analysis suggests
that the SDW at wave vectors $(2Q, 2Q)$ and $(2Q, -2Q)$ is
competing with CDW/VBS order parameters at $(2Q, 0)$ and $(0,
2Q)$, and the suppression of the SDW leads to the emerging of
CDW/VBS order parameters.

\section{Conclusions}
\label{sec:conc}

This paper has addressed a problem of long-standing interest
in the study of correlated electron systems in two spatial dimensions.
Many such systems have insulating, metallic, or superconducting
ground states with long-range antiferromagnetic order. By tuning the
electron concentration, pressure, or the values of exchange constants
in model systems, it is possible to drive a quantum phase transition to a
phase where the antiferromagnetic order is lost. We are interested in the
nature of the ``quantum-disordered'' phase so obtained.

For certain insulating square or honeycomb lattice models, the essential features were understood
some time ago \cite{haldane,rs1}: the lattice spins endow point spacetime defects
in the N\'eel order (`hedgehogs') with geometric (or Berry) phases, which lead to valence
bond solid (VBS) order in the quantum-disordered phase. Here we have presented a
more general version of this argument, in principle applicable to arbitrary insulating, metallic,
or superconducting electronic systems
in two dimensions, with general band structures. The key step was to associate
the geometric phases with bands of one electron states in the background of local antiferromagnetic
order. The antiferromagnetic order was then allowed to have a spacetime variation
in orientation (but {\em not\/} in magnitude) so that there was no long-range antiferromagnetic order,
thus accessing the quantum-disordered phase.
We found that the skyrmion density in this local antiferromagnetic order induced a response
in an electronic bilinear conjugate to the competing order:
this is contained in our key result in Eq.~(\ref{omain}).

Our main application of these results was to cases in which the electronic band structure
was fully gapped in the phase with antiferromagnetic order: we considered square lattice insulators
in Section~\ref{sec:neel}, honeycomb lattice insulators in Section~\ref{sec:honeycomb}, and $d$-wave
superconductors with spin density wave order nesting the nodal points in Section~\ref{sec:nodal}.
We obtained VBS order in many cases, but also found a number of other possible orderings.

However, in principle, the result Eq.~(\ref{omain}) applies also in cases where the antiferromagnetic
order does not fully gap the electron bands {\em e.g.\/} when there are hole and/or electron pockets.
Such a situation is clearly of importance for the underdoped cuprate superconductors.
The result in Eq.~(\ref{omain}) contains a singular dependence on $\bk$ at the Fermi surfaces
of such band structures, and this is likely of importance in the quantum-disordered phase.
Alternatively, expressions for the coupling $K$ in Section~\ref{sec:liang} would acquire long-ranged
corrections due to Fermi surface singularities.
We leave the elucidation of such effects to future work.
However, if we ignore such effects, the arguments of Section~\ref{sec:dual} would apply
also to this metallic case, with a variable exponent $\nu$ relating the monopole operator to the VBS order.
The net result is that any ordering associated with an integer power of $V$ is possible. Interestingly
the same conclusion was reached in an earlier study \cite{ribhu} of quantum disordered N\'eel states in
a compressible background using a toy model of bosons.

\acknowledgements

We thank T.~Grover and D.-H.~Lee for useful discussions. This
research was supported by the National Science Foundation under
grant DMR-0757145 and by a MURI grant
from AFOSR.

\appendix

\section{Rotor theory of Hubbard model}
\label{app:rotor}

This appendix will show how the decomposition in Eq.~(\ref{zpsi})
can be used to write an exact path integral representation of an
arbitrary Hubbard-like model. The $z_\alpha$ becomes co-ordinates
of an O(4) rotor in this path integral, and so do not directly
contribute to the geometric phases of interest in this paper. This
is to be contrasted from the alternative Schwinger boson
formulation, where the canonical nature of the Schwinger bosons
ensures that they carry the entire geometric phase at half-filling
\cite{rs2}.

We consider a Hubbard model on a general lattice \beq H = H_0 +
H_1 \eeq where $H_0$ has the single site terms \beq H_0 = \sum_i
\left[ U \left( n_{i \uparrow} - \frac{1}{2} \right) \left( n_{i
\downarrow} - \frac{1}{2} \right) - \mu (n_{i \uparrow} + n_{i
\downarrow} ) \right] \eeq and $H_1$ is the hopping term \beq H_1
= - \sum_{i<j} t_{ij} c_{i \alpha}^\dagger c_{j \alpha} \eeq

As in Eq.~(\ref{zpsi}), we transform the electron to a rotating
reference frame expressed in terms of the spinless fermions
$c_p$ and the complex unit spinor $z_\alpha$. Here, it is
useful to write $z_\alpha$ in real and imaginary parts: \beq
z_\uparrow = \phi_0 + i \phi_1 \quad, \quad z_\downarrow = \phi_2
+ i \phi_3 . \eeq The inner product of two complex spinors is \beq
\tilde{z}^\ast_\alpha z_\alpha = \tilde{\phi} (1 - \rho^y) \phi
\label{inner} \eeq We will use $\sigma^a$ for Pauli matrices in
the $\uparrow$, $\downarrow$ space, and $\rho^a$ for Pauli
matrices in the real/imaginary space. The global spin rotation
\beq z \rightarrow \left(1  + i \theta^a \sigma^a \right)
\label{zspin} z \eeq acts on $\phi$ via \beq \phi \rightarrow
\left(1 + i \theta^a S^a \right) \phi, \eeq where $S^a$ are the
antisymmetric Hermitian matrices \beq S^x = -\sigma^x \rho^y
\quad, \quad S^y = \sigma^y \quad, \quad S^z = -\sigma^z \rho^y .
\eeq Combining (\ref{inner}) and (\ref{zspin}) we have \beq z^\ast
\sigma^a z = \phi ( 1- \rho^y) S^a \phi = - \phi \rho^y S^a \phi.
\eeq The SU(2) gauge rotation \cite{su2} acts on $\psi$ as \beq
\psi \rightarrow (1 + i \theta^a \widetilde{\sigma}^a)
\label{psigauge} \psi \eeq  where $\widetilde{\sigma}^a$ are Pauli
matrices in the $\pm$ space. This gauge rotation acts on $z$ as
\beq \phi_\ell \rightarrow \left(1 + i \theta^a T^a_{\ell m}
\right) \phi_m, \label{phigauge} \eeq where the indices $\ell, m =
1 \ldots 4$ and $T^a$ are the antisymmetric Hermitian matrices
\beq T^x = \sigma^y \rho^x \quad, \quad T^y = - \sigma^y \rho^z
\quad, \quad T^z = \rho^y . \eeq

The physical states on a single site, which are eigenstates of
$H_0$, are \beqn c^\dagger_{\alpha} |0 \rangle & \leftrightarrow &
\left( z_{\alpha}^\ast \psi_+^\dagger - \epsilon_{\alpha\beta}
z_{\beta} \psi_-^{\dagger} \right) |0 \rangle \nn |0 \rangle &
\leftrightarrow &|0 \rangle \nn c^\dagger_{\uparrow}
c^{\dagger}_{\downarrow} |0 \rangle & \leftrightarrow &
\psi_+^\dagger \psi_-^\dagger |0 \rangle \label{eq:states} \eeqn
These 4 states have energies $-\mu - U/4$, $-\mu - U/4$, $U/4$,
and $U/4 - 2 \mu$.

Following Hermele \cite{hermele}, let us write these states in a
different manner, using the energy levels of a O(4) quantum rotor.
All of the following will work on a single site, and so we will
drop the site index. We will equate the states of $\phi$ to that
of a quantum particle moving on $S^3$ with co-ordinate $\phi$. On
this space, we introduce the angular momentum operators \beq
\mathcal{S}^a = -i \phi_\ell S^a_{\ell m} \frac{\partial}{\partial
\phi_m} \quad , \quad \mathcal{T}^a = -i \phi_\ell T^a_{\ell m}
\frac{\partial}{\partial \phi_m}. \eeq In the fermion sector we
have the usual angular momentum \beq \mathcal{L}^a =
\psi^\dagger_{p} \widetilde{\sigma}^a_{pp'} \psi_{p'} \label{lpsi}
\eeq Then all the states in Eq.~(\ref{eq:states}) satisfy
\begin{equation}
\mathcal{T}^a + \mathcal{L}^a = 0.
\label{eq:const}
\end{equation}

Now consider the following Hamiltonian for the rotor and the
fermions \beq \mathcal{H}_0 = K_1 \mathcal{S}^{a2} + K_2
\mathcal{T}^{a 2} + K_3 \psi_p^\dagger \psi_p + K_4 \psi_+^\dagger
\psi_-^\dagger \psi_- \psi_+ + K_5 \label{eq:rotor} \eeq For
appropriate ranges of the $K_i$ couplings, the low-lying states of
this Hamiltonian which obey Eq.~(\ref{eq:const}) map onto the
states of the $H_0$. The zero rotor-angular momentum states must
have 0 or 2 fermions, and these map onto the lower two states in
Eq.~(\ref{eq:states}), yielding \beqn K_5 &=& U/4 \nn 2 K_3 +K_4 +
K_5 &=& U/4 - 2 \mu \eeqn There are 4 rotor states with angular
momentum 1 and wavefunction $\sim \phi_\ell / |\phi|$. Because of
the constraint in Eq.~(\ref{eq:const}), these states must be
paired with states with fermion number 1. There are 2 such states,
leading to a total of 8 states. However, the conditions in
Eq.~(\ref{eq:const}) eliminate 6 of these states (there are 3
constraints for each fermion polarization), and so only 2 states
remain, as in the Hubbard model. The energy of these states yields
\beq 3K_1 + 3 K_2 + K_3 + K_5 = -\mu - U/4. \eeq The $K_i$
constants are over-determined, and in an exact treatment of the
constraint in Eq.~(\ref{eq:const}), the precise choice will not
matter. Of course, in mean-field theory, different choices will
lead to somewhat different results.

Now, following Hermele \cite{hermele}, we can write
Eq.~(\ref{eq:rotor}) as a path integral over $\phi_\ell (\tau)$
and $\psi_p (\tau)$ and obtain the Lagrangian \beqn \mathcal{L}_0
&=& \frac{1}{4 (K_1 + K_2)} \left[ \left( \partial_\tau \phi_\ell
- i A_\tau^a T^a_{\ell m} \phi_m \right)^2
 + \Delta^2 \phi_m^2 \right]
\nn &~&~+ \psi_p^\dagger \left( \partial_\tau \delta_{pp'} - i
A_\tau^a \widetilde{\sigma}^a_{pp'} \right) \psi_{p'}  + K_3
\psi_p^\dagger \psi_p + K_4 \psi_+^\dagger \psi_-^\dagger \psi_-
\psi_+ \eeqn where $A^a_\tau$ is the time-component of a SU(2)
gauge field which imposes the constraint (\ref{eq:const}), and
$\Delta^2$ imposes the unit length constraint on $z_\alpha$. We
can also insert the parameterization (\ref{zpsi}) into $H_1$ and
obtain the Lagrangain
\begin{eqnarray}
&& \mathcal{L}_1 = - \sum_{i<j} t_{ij} \Biggl[   \bigl( z_{i \alpha}^\ast z_{j \alpha} \bigr)
\left( \psi^\dagger_{i+} \psi_{j+} + \psi^\dagger_{j-} \psi_{i-} \right)
\nonumber \\
&&~~~~~~~~~+ \bigl( z_{j \alpha}^\ast z_{i \alpha} \bigr)
\left( \psi^\dagger_{i-} \psi_{j-} + \psi^\dagger_{j+} \psi_{i+} \right) \nonumber \\
&&~~~~~~~~~+\bigl( \varepsilon^{\alpha\beta} z_{j \alpha}^\ast z_{i \beta}^\ast \bigr)
\left( \psi^\dagger_{i+} \psi_{j-}
- \psi_{j+}^\dagger \psi_{i-} \right) \nonumber \\
&&~~~~~~~~~+ \bigl( \varepsilon^{\alpha\beta} z_{i \alpha} z_{j
\beta} \bigr) \left( \psi^\dagger_{i-} \psi_{j+} -
\psi^\dagger_{j-} \psi_{i+} \right) \Biggr] \label{hop}
\end{eqnarray}

There is now a natural mean field theory of $\mathcal{L}_0  +
\mathcal{L}_1$ which should yield all 4 phases of
Ref.~\onlinecite{su2}. The approximations are:
\begin{itemize}
\item Ignore the gauge field $A^a_\tau$. \item Factorize the
4-Fermi term, $K_4$ into $N^a \psi_p^\dagger
\widetilde{\sigma}^a_{pp'} \psi_{p'}$ The field $N^a$ is to be
determined self-consistently, and will be site-dependent. \item
Factorize $\mathcal{L}_1$ into fermion and boson bilinears, as
indicated by the parentheses. \item Phases A and C will also have
a $\phi$ condensate. It should be sufficient to work with $\langle
\phi \rangle =0$ in phases B and D, and determine their boundaries
to phases A and C \item Phase D should have $N^a = 0$, and also
$\langle z^\ast_\alpha z^\ast_\beta \rangle =0$ and $\langle
\psi_+^\dagger \psi_- \rangle = 0$. \item The value of $\Delta^2$
is determined as usual by solving the unit length constraint on
$z_\alpha$.
\end{itemize}

\section{Square lattice antiferromagnetic in an applied gauge flux}
\label{app:gauge}

This appendix will carry out a computation similar to that
of Section~\ref{sec:linear} using gauge-theoretical formulation in Eq.~(\ref{e3}).
Rather than a slowly varying N\'eel order $n^a (\br)$ as in Eq.~(\ref{e2}), this appendix
will have a slowly varying gauge potential ${\bf A} (\br )$. The results here will
be connected to those of Section~\ref{sec:linear} via Eq.~(\ref{gaugeskyrmion}).
However, a precise quantitative equivalence between Eqs.~(\ref{e2}) and (\ref{e3}) requires
inclusion of the last two terms of Eq.~(\ref{hop}) in Eq.~(\ref{e3}), which we will not
account for here. The importance of these omitted terms should be clear from Appendix A
of Ref.~\onlinecite{yang}.

Now we expand Eq.~(\ref{e3}) to first order in $A_{ij}$, and using
Eq.~(\ref{Aij}) we can write $H=H_0+H_1$ where
$H_0$ has the same form as Eq.~(\ref{eh0}) but with the $\psi_\pm$ fermions
\beqn H_0 &=& \sum_{{\bk}} \left( \varepsilon_{{\bk}}
\psi^{\dagger} ({\bk}) \psi ({\bk}) + m \psi^\dagger ({\bk} +
\bQ) \sigma^z \psi ({\bk}) \right),
\label{eh0psi}
\eeqn
while $H_1$ in Eq.~(\ref{a2}) is replaced by
\beqn H_1 &=& - i \sum_{i<j} t( {\bf
r}_i - {\bf r}_j) A_{ij} \left( \psi^\dagger_i \sigma^z \psi_j -
\psi^\dagger_j \sigma^z \psi_i \right) \nn &=& \sum_{{\bk},{\bq}}
\left[ {\bf A} ({\bq}) \cdot \frac{\partial
\varepsilon_{\bk}}{\partial {\bk}} \right] \psi^\dagger ({\bk} +
{\bq}/2) \sigma^z \psi ({\bk} - {\bq}/2)  + \mathcal{O} ({\bq}^2)
\label{e1a}
\eeqn
Note that $H_1$ does not include the omitted terms represented by the ellipses
in Eq.~(\ref{e3}), which appear as the last two terms in Eq.~(\ref{hop}); this will
be significant below.

Now we will use the Kubo formula to determine the response to the
applied gauge field in $H_1$. We will work to linear response order ${\bf
A}$, and to linear order in ${\bf q}$.

We have to carefully define an observable: it should be gauge
invariant and spin-rotation invariant. For this reason we look at
the response in the following \beq M_{ij} \equiv \psi^\dagger_i
e^{i \sigma^z A_{ij}} \psi_j \label{Mij} \eeq We want to compute
the change in $\langle M_{ij} \rangle$ to linear order in ${\bf A}
({\bq})$, and in the limit of small ${\bq}$. In momentum space
\beqn
\langle M_{ij} \rangle &=& \sum_{{\bk}, {\bp}}  e^{- i \bk
\cdot {\bf r}_i + i \bp \cdot {\bf r}_j}
\left\langle \psi^{\dagger} (\bk) \,;\,\psi (\bp) \right\rangle \label{dmij} \\
&~&+ i \left[\sum_{{\bq}} {\bf A} ({\bq}) \cdot ({\bf r}_j - {\bf
r}_i ) \, e^{i {\bq} \cdot ({\bf r}_j+ {\bf r}_i)/2} \right]
\left[ \sum_{\bk}  e^{- i \bk \cdot ({\bf r}_i - {\bf r}_j)} e^{i
\bQ \cdot {\bf r}_j} \left \langle \psi^\dagger (\bk) \sigma^z
\psi (\bk + \bQ) \right\rangle \right] \nonumber
\eeqn
In the
second term, we have assumed we are expanding to linear order in
${\bf A}$, and so assumed momentum conservation in the fermion
bilinear expectation value.

We have to expand the first term to linear order in ${\bf A}$, and
so we expand the second term in Eq.~(\ref{dmij}) using Wick's theorem.
\beqn && \left\langle \psi^\dagger (\bk ') \psi (\bp) \,;\,
\psi^\dagger ({\bk} + {\bq}/2) \sigma^z \psi ({\bk} - {\bq}/2)
\right \rangle  \nn && = - 2 \sum_\omega \Biggl[ \delta_{\bp, \bk
+ \bQ + \bq/2}  \delta_{\bk',\bk - \bq/2} G(\bk - \bq/2) F (\bk + \bq/2)  \nn
&&~~~~~~~~~~~~~~~~~+ \delta_{\bp, \bk + \bq/2}  \delta_{\bk',\bk  +\bQ -
\bq/2} G (\bk + \bq/2) F (\bk - \bq/2)
 \Biggr].
\eeqn
Also from Eq.~(\ref{green})
\beq
\left \langle \psi^\dagger (\bk) \sigma^z \psi
(\bk + \bQ) \right\rangle =  -2 \sum_\omega F(\bk)
 \eeq

Putting everything together
\beqn
&& \delta \langle M_{ij} \rangle
=  2\sum_{{\bk}, {\bq},\omega} e^{- i \bk \cdot ({\bf r}_i - {\bf
r}_j)} e^{i {\bq} \cdot ({\bf r}_j+ {\bf r}_i)/2}  e^{i \bQ \cdot
{\bf r}_j} {\bf A} ({\bf q}) \cdot \Biggl[ \nn &&~  \frac{\partial
\varepsilon_\bk}{\partial \bk} G(\bk - \bq/2) F (\bk + \bq/2)
+
  \frac{\partial \varepsilon_{\bk + \bQ}}{\partial \bk}
  G(\bk +\bQ + \bq/2) F (\bk - \bq/2)
 +
\frac{\partial F(\bk) }{\partial {\bf k}} \Biggr]
 \label{e4}
\eeqn
Explicit evaluation shows that the expression in the square
brackets does indeed vanish at ${\bf q} = 0$, as is required by
gauge invariance. Now expand Eq.~(\ref{e4}) to first order in $\bq$ and find
\beq
\delta \langle M_{ij}
\rangle = 2\sum_{{\bk}, {\bq}} e^{- i \bk \cdot ({\bf r}_i - {\bf
r}_j)} e^{i {\bq} \cdot ({\bf r}_j+ {\bf r}_i)/2}  e^{i \bQ \cdot
{\bf r}_j}  {\bf A} ({\bf q}) \cdot {\bf I} (\bk, \bq) \label{e11}
\eeq
where
\beq
{\bf I} (\bk, \bq) =
\left[\frac{\partial \varepsilon_\bk}{\partial \bk} \left( \bq
\cdot \frac{\partial \varepsilon_{\bk+\bQ}}{\partial \bk} \right)
- \frac{\partial \varepsilon_{\bk+\bQ}}{\partial \bk} \left( \bq
\cdot \frac{\partial \varepsilon_\bk}{\partial \bk}
 \right) \right] \sum_\omega  \frac{m/2}{(-i \omega + E_{1 \bk})^2 (-i \omega  + E_{2 \bk})^2}
 \label{e12}
 \eeq
Combining (\ref{e11}) and (\ref{e12}), we have the result analogous to Eq.~(\ref{omain}):
\beq
\left \langle c^\dagger (\bk) c (\bk + \bQ) \right\rangle =
-2 i \widetilde{\mathcal{F}} (\bk)  \left (\partial_x A_y - \partial_y A_x
\right) \label{main}
\eeq
where
\beqn
\widetilde{\mathcal{F}} (\bk) &=&
\left( \frac{\partial
\varepsilon_{\bk+\bQ}}{\partial \bk} \times \frac{\partial
\varepsilon_\bk}{\partial \bk} \right)
\sum_\omega  \frac{m/4}{(-i \omega + E_{1 \bk})^2 (-i \omega  + E_{2 \bk})^2} \nn
&=& \frac{m}{2} \left( \frac{\partial
\varepsilon_{\bk+\bQ}}{\partial \bk} \times \frac{\partial
\varepsilon_\bk}{\partial \bk} \right)
\frac{( \mbox{sgn} (E_{1\bk}) - \mbox{sgn}(E_{2 \bk}))}{(E_{1 \bk} - E_{2 \bk})^3} .
\label{f}
\eeqn
We have
written Eq.~(\ref{main}) in terms of the original electron
operators $c (\bk )$: we are working to linear order in ${\bf A}$,
and so this order all variables can be mapped onto the original
gauge-invariant operators. Comparing Eq.~(\ref{f}) with Eq.~(\ref{J3}), and using Eq.~(\ref{gaugeskyrmion}), we
should expect equality between $\mathcal{F} (\bk )$ and $\widetilde{\mathcal{F}} (\bk)$. However, while both
functions have an identical symmetry structure, and similar singularities at possible Fermi surfaces (which is all we need), they are not precisely equal.
This can be traced to the absence of precise equality between Eqs.~(\ref{e2}) and (\ref{e3}), due to the omission
of the last two terms in Eq.~(\ref{hop}), which were also important in previous computations \cite{yang}.

\end{document}